\documentstyle[12pt,epsf]{article}
\def\be{\begin{equation}}
\def\ee{\end{equation}}

\def\be{\begin{equation}}
\def\ee{\end{equation}}
\begin{document}

\begin{center}

{ Preprint Institute for basic Research IBR-TP-12-05}

{December 15, 2005, submitted for publication}

\vspace*{1.00cm}

{\large \bf LIE-ADMISSIBLE INVARIANT ORIGIN OF\\ IRREVERSIBILITY FOR MATTER AND ANTIMATTER\\
AT THE CLASSICAL AND OPERATOR LEVELS}

\vspace{0.50cm}
{\bf  Ruggero Maria  Santilli}\\  {Institute for  Basic Research}\\
{P. O. Box  1577, Palm Harbor,  FL 34682,  U.S.A.}\\
{ibr@gte.net, http://www.i-b-r.org}

\end{center}

\begin{abstract}

It was generally  believed throughout the 20-th century
that irreversibility is a purely classical event without
operator counterpart. However, a classical irreversible
system cannot be consistently decomposed into a finite
number of reversible quantum particles (and, vice versa), thus
establishing that the origin of irreversibility is
basically unknown at the dawn of the 21-th century. To
resolve this problem, we adopt the historical
analytic representation of irreversibility by Lagrange and Hamilton
with external terms in their analytic equations; we show that,
when properly written, the brackets of the time evolution
characterize covering Lie-admissible algebras; we
show that the formalism has a fully consistent operator
counterpart given by the Lie-admissible branch of
hadronic mechanics; we identify catastrophic mathematical and
physical inconsistencies when irreversible formulations are
treated with the conventional mathematics used for reversible
systems; and show that, when the
dynamical equations are treated with a novel irreversible
mathematics, Lie-admissible formulations are fully consistent
because invariant at both the classical and operator
level. The case of  closed-isolated systems verifying conventional total
conservation laws, yet possessing an irreversible structure, is
treated via the Lie-isotopic subclass of Lie-admissible
formulations. The analysis is conducted for both
matter and antimatter at the classical and operator levels to
prevent insidious inconskistenmcies occurring for the sole study
of matter or, separately, antimatter, as known known for the
problem of grand unificationsr.

\end{abstract}
\vskip0.30cm

\noindent {\bf 1. INTRODUCTION}

\noindent {\bf 1.1 The Scientific Imbalance Caused by Irreversibility.}
As it is well known, rather vast studies have been conducted in the 20-th century
in attempting a reconciliation of the manifest irreversibility in time of our
macroscopic environment with the reversibility of quantum mechanics, resulting
in the rather widespread belief that irreversibility solely exists at the
macroscopic level because it ''disappears'' at quantum mechanical level (see, e.g.,
the reprint volume by Shoeber and vast literature quoted therein).

The above belief created a large scientific imbalance because irreversible
systems were treated  with   the
mathematical and physical formulations developed for
 reversible systems. Since these formulations are themselves
reversible, the attempt in salvaging quantum mechanics {\it vis a vis} the
irreversibility of physical reality caused  serious limitations
in virtually all quantitative sciences.

The imbalance originated from the fact that all used formulations
were of {\it Hamiltonian type} (i.e., the formulations are entirely characterized
by  the sole knowledge of the Hamiltonian), under the awareness that all known
Hamiltonians are reversible because all known potentials (such as the Coulomb
potential $V(r) =  {q_1q_2\over r}$) are  reversible).

The academic belief of the purely classical character of irreversibility
 was disproved in 1985 by Santilli [2] via the following
theorem whose proof is instructive for serious scholars in the field
\vskip0.30cm

{\it THEOREM 1.1: A classical irreversible system cannot be consistently reduced
to a finite number of quantum particles all in reversible conditions and, vice
versa, a finite ensemble of quantum particles all in reversible conditions cannot
consistently reproduce an irreversible macroscopic system under the
correspondence or other principles.}

\vskip0.30cm

The implications of the above theorem are rather deep because the theorem
establishes that, rather than adapting to quantum mechanics all possible physical
events in the universe, it is necessary to seek a
{\it covering of quantum mechanics} permitting a consistent
treatment of elementary particles in irreversible conditions, where the notion of
covering is intended to be such that, when irreversibility is removed, quantum
mechanics is recovered identically and uniquely.

In the final analysis,
the orbit of an electron in an atomic structure is indeed reversible in time, with
consequential validity of quantum mechanics. However, the idea that the
same electron has an equally reversible orbit when moving in the core of a star is
repugnant to reason as well as scientific vigour because, e.g., said idea
would imply that the electron has to orbit in the core of a star
with a locally conserved angular momentum (from the basic rotational symmetry of
all Hamiltonians), with consequential direct belief of the existence of the
perpetual motion within physical media.

In short, contrary to a popular belief of the 20-th century, Theorem 1.1
establishes that  {\it irreversibility originates at the most primitive levels of
nature, that of elementary particles, and then propagates all the way to our
macroscopic environment.}

In this paper, we complete studies in the origin of irreversibility initiated by
the author during his graduate studies in physics at the University of Torino,
Italy, in the late 1960s [7-9], and continued at Harvard University under DOE support in
the late l970s [6,11,12], which studies achieved mathematical and physical consistency
only recently. The decades required by the completion of the research is an indication
of the complexity of the problem.

In fact, it was relativity easy to identify since the early efforts
irreversible generalizations of Heisenberg's equations [7-9]. However, these
generalized equations subsequently resulted to lack {\it invariance,} here
intended as the capability by quantum mechanics of predicting the same numerical
values for the same conditions but under the time evolution of the theory.

In turn, the achievement of invariance
predictably required the laborious prior effort of identifying a {\it new
irreversible mathematics,} namely, a new mathematics specifically constructed for
the invariant treatment of irreversible systems while being a covering of
conventional mathematics in the above indicated sense. Following the achievement of the
new mathematics, the resolution of the scientific imbalance of the 20-th century
caused by irreversibility was direct and immediate.

By looking in retrospective, rather than  being demeaning for quantum mechanics,
the search for its invariant irreversible covering has brought into full light
the majestic axiomatic consistency of quantum mechanics, and how difficult
resulted to be the preservation of the same axiomatic consistency at the
covering irreversible level.

In other words,  only art masterpieces, such as
Michelangelo's {\it La Piet\'a,} are eternal. By comparison, quantitative
sciences will never admit final theories because, no matter how majestic a given
theory may appear at a given time, its surpassing by a broader theory for more
complex physical conditions is only a matter of time.

As we shall see, the studies presented in this paper required several preceding contributions over an
extended period of time, although their coordination for the invariant treatment of irreversibility is
presented in this paper for the first time.  A comprehensive presentation of these studies
will appear in monographs [18c] following the publication of this paper.

\vskip0.30cm


\noindent {\bf 1.2 The Forgotten Legacy of Newton, Lagrange and Hamilton.}
The scientific imbalance on irreversibility was created in
the early part of the 20-th
century when, to achieve compatibility  with quantum mechanics and
special relativity, the entire universe was reduced to potential forces and the analytic
equations  were ``truncated" with the removal of the external terms.

In reality, Newton [3] {\it did not} propose his celebrated equations to be
restricted to reversible forces derivable from a potential
$F = \partial V / \partial r$, but proposed them for the most general
possible irreversible systems,
$$
m_a\times \frac {dv_{ka}}{dt} = F_{ka}(t, r, v),\; \; \; k = 1, 2, 3;
\; \; \; a = 1, 2, \ldots, N,
\eqno(1.1)
$$
where the conventional associative product of numbers, matrices, operators, etc.,
is  denoted hereon with the symbol
$\times$ so as to distinguish it from numerous other products needed later on.

 Similarly, to be compatible with Newton's equations,
Lagrange [4] and Hamilton [5] decomposed Newton's force into a potential and a
nonpotential component, represented all
potential forces with functions today known as the Lagrangian and the Hamiltonian,
and represented nonpotential forces with external terms. Therefore the {\it true
Lagrange and Hamilton equations} are given, respectively, by
$$
\frac{d}{dt}\frac{\partial L(t, r, v)}{\partial v_a^k} - \frac{\partial
L(t, r, v)}{\partial r_a^k} =  F_{ak}(t, r, v),
\eqno(1.2a)
$$
$$
\frac{dr_a^k}{dt} = \frac{\partial H(t, r, p)}{\partial p_{ak}}, \; \; \;
\frac{dp_{ak}}{dt} = -
\frac{\partial H(t, r, p)}{\partial r_a^k} + F_{ak}(t, r, p),
\eqno(1.2b)
$$
$$
L = {\large \Sigma}_a\frac{1}{2} \times m_a\times { v}_a^2 - V(t, r, v), \; \; \;
 H = {\large \Sigma}_a\frac{{\bf p}_a^2}{2\times m_a} + V(t, r, p),
\eqno(4.1.2c)
$$
$$ V =  U(t, r)_{ak}\times v_a^k + U_o(t, r), \; \; \;
F(t, r, v) = F(t, r, p/m).
\eqno(1.2d)
$$
The analytic equations used throughtout the 20-th century, those without
external terms, shall be referred to as the {\it truncated Lagrange and Hamilton equations.}

 More recently, Santilli [6a] conducted comprehensive studies on the integrability
conditions for the existence of a
potential, or a Lagrangian, or a Hamiltonian, called {\it conditions of variational
selfadjointness.} These study permit
the rigorous decomposition of Newtonian forces into a component that is
variationally selfadjoint (SA) and a component
that is not (NSA),
$$
m_a\times \frac {dv_{ka}}{dt} = F^{SA}_{ka}(t, r, v) + F^{NSA}_{ka}(t, r, v).
\eqno(1.3)
$$

Consequently, the true Lagrange and Hamilton equations can be more technically written
$$
\Bigl[\frac{d}{dt}\frac{\partial L(t, r, v)}{\partial v_a^k} - \frac{\partial
L(t, r, v)}{\partial r_a^k}\Bigr]^{SA} =  F^{NSA}_{ak}(t, r, v),
\eqno(1.4a)
$$
$$
\Bigl[\frac{dr_a^k}{dt} - \frac{\partial H(t, r, p)}{\partial p_{ak}}\Bigr]^{SA} = 0, \; \; \;
\Bigl[\frac{dp_{ak}}{dt} +
\frac{\partial H(t, r, p)}{\partial r_a^k}\Bigr]^{SA} = F^{NSA}_{ak}(t, r, p),
\eqno(1.4b)
$$

 The {\it forgotten  legacy
of Newton, Lagrange and Hamilton
is that irreversibility originates
precisely in the truncated  external  terms, } because all known potential-SA forces are
reversible. The scientific imbalance of
Section 1.1 is then due to the fact that no serious scientific study on irreversibility can
 be conducted with the
truncated analytic equations and their operator counterpart, since these equations
can only represent reversible
systems.

Being born and educated in Italy, during his graduate studies the author had the opportunity of
reading in the late 1960s the original works by Lagrange that were written  in Torino (where this
paper has beeen written) and mostly in Italian.

In this way, the author had the opportunity of verifying {\it Lagrange's analytic vision of
representing irreversibility precisely via the external terms,} due to the impossibility of
representing all possible physical events via the sole use of the Lagrangian, since the latter
was  conceived for the representation of reversible and potential events. As the reader
can verify, Hamilton had, independently, the same vision.

Consequently, the truncation of the analytic equations caused the impossibility of a
credible treatment of irreversibility at the purely classical level. The lack of a credible
treatment of irreversibility then propagated at the subsequent quantum mechanical and quantum field
theoretical levels.

It then follows that {\it quantum mechanics cannot possibly be used for serious
studies on irreversibility} because the
discipline was constructed for the description of reversible quantized atomic orbits
and not for irreversible systems.

While the validity of quantum mechanics for the arena of its original
conception and construction is
beyond scientific doubt, the assumption of quantum mechanics as the final operator
theory for all conditions existing in
the universe, such as orbits of particles in the core of a star, is outside the boundaries of
serious science.

This establishes  the need for the construction of a
covering of quantum mechanics specifically conceived and constructed for quantitative
treatments of irreversible systems.

\vskip0.30cm

\noindent {\bf 1.3 Early Representations of Irreversible Systems.}
As it is well known, the brackets of the time evolution of the truncated analytic equations,
the familiar {\it Poisson brackets,} characterize a {\it Lie algebra,} a feature that persists
at the quantum level, thus establishing Lie algebras as the ultimate foundations of the physics
of the 20-th century.

By contrast, the brackets of the time evolution of
an observable $A(r, p)$ in phase space according to the analytic
equations with external terms,
$$
 {dA\over dt} = (A, H, F) =  \frac{\partial A}{\partial r_a^k}\times
\frac{\partial H}{\partial p_{ka}} - \frac{\partial H}{\partial
r_a^k}\times \frac{\partial A}{\partial p_{ka}} + {\partial A\over
\partial r^k_a}\times F_{ka},
\eqno(1.5)
$$
 cannot characterize any {\it algebra} as commonly understood in
mathematics, because the brackets violate the right associative and scalar laws.

Therefore, the presence of external terms in the analytic equations
causes not only the loss of {\it all} Lie algebras in the study of
irreversibility, but actually the loss of all possible algebras.

To resolve this problem, the author initiated a long scientific
journey  following the reading of Lagrange's papers.

The original argument [7-9], still valid today, is to select analytic
equations characterizing brackets in the time evolution that verify
the following conditions:

(1) Said brackets must verify the right and
left associative and scalar laws to characterize an algebra;

(2) Said brackets must not be invariant under time reversal as a
necessary condition to represent irreversibility {\it ab initio;} and

(3) The underlying algebras must be a covering of Lie algebras as a necessary
condition to characterize a covering of the truncated analytic equations,
namely, as a condition for the selected representation of
irreversibility to admit reversibility as a particular case.

Condition (1) requires that said brackets should be bilinear, e.g., of
the form $(A, B)$ with basic algebraic axioms
$$
(n\times A, B) = n\times (A, B),\; \; \; (A, m\times B) = m\times
(A, B); \ \ n, m \in C,
\eqno(1.6a)
$$
$$
(A\times B, C) = A\times (B, C), \; \; \; (A, B\times C) = (A,
B)\times C.
\eqno(1.6b)
$$
Condition (2) can be first realized by requiring that brackets $(A, B)$ are not totally
antisymmetric as the conventional Poisson brackets,
$$
(A, B) \not = - (B, A),
\eqno(1.7)
$$
because time reversal is realized via the use of Hermitean
conjugation.

 Condition (3)  implies that  brackets $(A, B)$
characterize  {\it Lie-ad\-mis\-sible algebras} in the  sense of Albert
[10], with operator form resulting to be also Jordan-admissible according to the following

\vskip0.30cm

{\it DEFINITION 1.1: A generally nonassociative
algebra
$U$ with elements $a, b, c, \ldots$ and abstract product $ab$ is said
to be Lie-admissible when the attached algebra $U^-$ characterized by the same vector space $U$
equipped with  the product $[a, b] = ab - ba$ verifies the  Lie axioms}
$$
[a, b] = - [b, a],
\eqno(1.8a)
$$
$$
[[a,  b], c] + [[b, c], a] + [[c, b], a] = 0.
\eqno(1.8b)
$$
{\it Said  generally nonassociative algebra $U$  is said to be
Jordan-admissible when the attached algebra $U^+$ characterized by the vector space U
equipped with the product $\{a, b\} = ab + ba$ verifies the  Jordan axioms}
$$
\{a, b\} =  \{b, a\},
\eqno(1.9a)
$$
$$
\{\{a, b\}, a^2\} = \{a, \{b, a^2\}\}.
\eqno(1.9b)
$$

In essence, the condition of Lie-admissibility requires that the brackets $(A,B)$ contain a
totally antisymmetric component with a residual totally symmetric part, thus admitting the
decomposition
$$
(A, B) = [A, B]^* + \{A, B\}^*.
\eqno(1.10)
$$
where the $^*$ denotes expected generalizationms of conventional brackets.

The antisymmetric brackets $[A, B]^*$ generally  result to be Lie at both classical and
operator levels. However, as illustrated below, for reasons yet unknown the  symmetric brackets
$\{A, B
\}^*$ always verify Jordan's first axiom (1.9a), but they generally violate the second axiom (1.9b)
in their classical form and verify it only in their operator form.

After identifying the above lines, Santilli [9] proposed in 1967
the following {\it generalized analytic equations}
$$
\frac{dr_a^k}{dt} = \alpha \times \frac{\partial H(t, r,
p)}{\partial p_{ak}}, \; \; \; \frac{dp_{ak}}{dt} = - \beta
\times\frac{\partial H(t, r, p)}{\partial r_a^k},
\eqno(1.11)
$$
(where $\alpha$ and $\beta$ are real non-null parameters) that are
manifestly irreversible. The time evolution is then
given by
$$
{dA\over dt} = (A, H) =
$$
$$
= \alpha \times \frac{\partial A}{\partial r_a^k}\times
\frac{\partial H}{\partial p_{ka}} -  \beta \times \frac{\partial
H}{\partial r_a^k}\times \frac{\partial A}{\partial p_{ka}},
\eqno(1.12)
$$
whose brackets are manifestly Lie-admissible because the attached antisymmetric brackets are
proportional to the conventional Poisson brackets,
$$
[A, B]^* = (A, B) - (B, A) = (\alpha + \beta) \times (\frac{\partial A}{\partial r_a^k}\times
\frac{\partial H}{\partial p_{ka}} -   \frac{\partial
H}{\partial r_a^k}\times \frac{\partial A}{\partial p_{ka}}).
\eqno(1.13)
$$

However, the attached symmetric brackets
$$
[A, B]^* = (A, B) + (B, A) = (\alpha - \beta) \times (\frac{\partial A}{\partial r_a^k}\times
\frac{\partial H}{\partial p_{ka}} +   \frac{\partial
H}{\partial r_a^k}\times \frac{\partial A}{\partial p_{ka}}),
\eqno(1.14)
$$
do not characterize a Jordan algebra because they verify first axiom (1.9a) but violate the
second axiom (1.9b) as the reader is encouraged to verify.

The above analytic equations characterize the time-rate of variation
of the energy
$$
{dH\over dt} = (\alpha - \beta)\times \frac{\partial H}{\partial
r_a^k}\times \frac{\partial H}{\partial p_{ka}}.
\eqno(1.15)
$$

Also in 1967, Santilli [7,8] proposed an operator counterpart of the
preceding classical setting consisting in the first known {\it
Lie-admissible parametric generalization of Heisenberg's equation}
in the following infinitesimal and finite forms
$$
i\times {dA\over dt} = (A, B) =  p\times A\times H - q\times H\times
A =
$$
$$
= m\times (A\times B - B\times A) + n\times (A\times B + B\times A),
\eqno(1.16a)
$$
$$
A(t) = W(t)\times A(0)\times W^{\dag}(t) = e^{i\times H\times q\times t}\times A(0)\times e^{-i\times
t\times p\times H},
\eqno(1.16b)
$$
$$
 W\times W^\dag \not = I.
\eqno(1.16c)
$$
where $p, q, p\pm q$ are non-null parameters, and
$$
m = p + q, \; \; \; n = q - p,
\eqno(1.17)
$$

Brackets $(A, B)$ are manifestly Lie-admissible with attached
antisymmetric part
$$
[A, B]^* = (A, B) - (B, A) = (p - q)\times [A, B].
\eqno(1.18)
$$
The same brackets are also Jordan-admissible as interested readers
are encouraged to verify, thus illustrating the peculiar occurrence whereby symmetric brackets
that are apparently similar are Jordan admissible only at the operator level.

Despite this limitation, the Jordan-admissibility of the operator brackets establishes Jordan's
dream on the physical applications of his algebras, although the latter occur for irreversible
systems. As a matter of fact, the above Jordan-admissibility establishes the impossibility for
Jordan algebras to have applications at the purely quAntum level, due to its purely reversible,
thus Lie character.

The  time evolution of Eqs. (1.16) is manifestly irreversible (for $p
\not = q$) with nonconservation of the energy
$$
i\times {dH\over dt} = (H, H) = (p - q)\times H\times H \not = 0.
\eqno(1.19)
$$

Subsequent to papers [7-9], Santilli realized that the above formulations are not
invariant under their own time evolution because Eq. (1.16b) is
manifestly {\it nonunitary.}

The application of nonunitary transforms to time evolution (1.16) then
led to the proposal in memoir [11,12] of 1978 of the following {\it
Lie-admissible operator generalization of Heisenberg equations} in
the following infinitesimal and finite forms
$$
i\times {dA\over dt} = A < H - H > A = A\times R\times H - H\times S\times A = (A,
H)^*,
\eqno(1.20a)
$$
$$
A(t) = W(t)\times A(0)\times W^\dag(t) = (e^{i\times H\times S\times t})\times A(0)\times
(e^{-i\times t\times R\times H}), \; \; \; W\times W^\dag \not = I,
\eqno(1.20b)
$$
$$
R = R(t, r, p, \psi, \partial \psi, ...) = S^\dag.
\eqno(1.20c)
$$
where $R$, $S$ and $R \pm S$ are now nonsingular operators (or
matrices), and Eq.~(1.20c) is a basic consistency condition
explained later on.

Eqs. (1.20) are the {\it fundamental equations of hadronic
mechanics} 12,18]. Their basic brackets are manifestly Lie-admissible and
Jordan admissible with structure
$$
(A, B)^* = A < B - B > A = A\times R\times B - B\times S\times A =
$$
$$
= (A\times T\times B - B\times T\times A) + (A\times Q\times B +
B\times Q\times A),
\eqno(1.21a)
$$
$$
T = R + S, \; \; \; T = S - R.
\eqno(1.21b)
$$

The generalized classical equations proposed in Refs. [11,12] jointly with Eqs. [1.20] are
given in unified phase space notation by
$$
{db^\mu\over dt} = S^{\mu\nu}(t, b)\times {\partial H(b)\over \partial b^\nu},
\eqno(1.22a)
$$
$$
b = (b^\mu) = (r^k, p_k), \; \; \; \mu = 1, 2, ..., 6\; \; \; k = 1, 2, 3,
\eqno(1.22b)
$$

where the tensor $S^{\mu\nu}$ is Lie-admissible (but not jointly Jordan admissible),
namely, the attached antisymmetric tensor
$$
\Omega^{\mu\nu} = S^{\mu\nu} - S^{\nu\mu},
\eqno(1.23)
$$
characterizes Birkhoff'a equations (see monograph [6b] for comprehensive studies), with
solution
$$
S_{\mu\nu} = {\partial R_\nu(t, b)\over \partial b^\mu}, \; \; \; S^{\mu\nu} =
[(S_{\alpha\beta})^{-1}]^{\mu\nu}.
\eqno(1.24)
$$
The canonical particular case is recovered for the  value $R = (R_\mu) = (p_k,
0)$ as the reader is encouraged to verify [6b,11].

It is easy to see that the application
of a nonunitary transform to
the parametric equations (1.16) leads to the operator
equations (1.20) and that the application of additional nonunitary
transforms preserves their
Lie-admissible and Jordan-admissible characters.

Consequently,
fundamental equations (1.20) are
``directly universal'' in the sense of admitting as particular cases all possible brackets
characterizing an algebra (universality) without the use of the transformation theory (direct
universality).
\vskip0.30cm


\noindent {\bf 1.4 Catastrophic Inconsistencies of Early Irreversible Formulations.}
Despite their direct universality, Eqs. (1.20) {\it are not invariant
under their own time evolution and, consequently, are afflicted by  catastrophic
inconsistencies} [29-37].

To clarify this crucial point, let us recall that the
capability by quantum mechanics to predict the same numerical values under the same conditions
at different times is given by the unitary character of Heisenberg's time evolution for
Hermitean Hamiltonians,
$$
A(t) = U(t)\times A(0)\times U^\dag = (e^{i\times H\times t})\times A(0)\times (e^{-i\times
t\times H}),
\eqno(1.25a)
$$
$$
U \times U^\dag = U^\dag\times U = I, \; \; \; H = H^\dag,
\eqno(1.25b)
$$
that characterizes the following type of invariance of the Lie brackets,
$$
U\times [A, B]\times U^\dag = A'\times B' - B'\times A' = [A', B'],
\eqno(1.26a)
$$
$$
A' = U\times A\times U^\dag, \;\; \; B' = U\times B\times U^\dag .
\eqno(1.26b)
$$
namely, an invariance specifically referred to the preservation of the
conventional associative product, $\times ' \equiv \times$.

For the case of the broader Lie-admissible equatsions (1.20) the situation is different because
the Hamiltonian remains Hermitean, but the time evolution is nonunitary, in which case the
application of the time evolution leads to the {\it new} brackets
$$
W\times (A, B)^*\times W^\dag = A' <' B' - B' >' A' = A'\times R'\times B' - B'\times S'\times
S'\times A' =
$$
$$
= A' \times (W^{\dag -1}\times R\times W^{-1})\times B' - B'\times (W^{\dag -1}\times R\times
W^{-1})\times A' = {(A', B')^*}',
\eqno(1.27a)
$$
$$
W\times W^\dag \not = I, \; \; \; H = H^\dag,
\eqno(1.27b)
$$
namely, the transformed product remains Lie-admissible due to its direct universality, but the product
itself is altered,
$R\rightarrow R' \not = R, S\rightarrow S' \not = S$.

As we shall see in Section 4, different values  of the $R$ and $S$ operators characterize {\it
different physical systems.} Consequently, the above change of values of the $R$ and $S$ operators
caused by the time evolution of the theory cannot preserve numerical predictions in time.

It is instructive to verify that essentially the same occurrences hold for the classical
Lie-admissible equations [11] due to the noncanonical character of the time evolution.

Without proof we, therefore, quote the following:
\vskip0.50cm

{\it THEOREM 1.2 }[29]: {\it Whether Lie or Lie-admissible, all classical theories possessing
noncanonical time evolutions are afflicted by catastrophic mathematical and physical inconsistencies.}
\vskip0.50cm

{\it THEOREM 1.3} [29]: {\it ):  All operator theories possessing a
nonunitary time evolution formulated on conventional Hilbert spaces ${\cal H}$ over conventional
fields of complex numbers $C$ are afflicted by catastrophic mathematical and physical
inconsistencies. In particular, said nonunitary theories

 1) do not possess invariant units of time, space, energy, {\it etc.},
thus lacking physically meaningful application to
measurements;

 2) do not conserve Hermiticity in time, thus lacking physically
meaningful observables;

 3) do not possess unique and invariant numerical predictions;

 4) generally violate probability and causality laws; and

 5) violate the basic axioms of Galileo's and Einstein's relativities.}

\vskip0.50cm

A comprehensive presentation of the above inconsistency theorems is available in Chapter 1 of
monograph [18c]. For the limited scope of this paper it is sufficient to
indicate that the mathematical inconsistencies are rather serious. All mathematical
formulations used in physics are based on fields that, in turn, are centrally dependent
on the basic unit. However, noncanonical or nonunitary transformations do not preserve the
basic unit by central assumption.

Hence, a given field at the initial time is no longer applicable at a subsequent time for
all noncanonical or nonunitary theories, due to the lack of time invariance of the basic unit. In
turn, the loss under the time evolution of the base field causes the catastrophic collapse of all
mathematical formulations.

For example, the formulation of a noncanonical classical theory on the Euclidean space $\cal {E}$
over the field of Real numbers $R$, or of  a nonunitary operator theory on a Hilbert space $\cal
H$ over the field of complex numbers $C$ has no mathematical consistency, again, because of the lack
of invariance of the basic unit of the fields (and, hence, of the field themselves) under the time
evolution.

Conventional mathematics can at best be used for the treatment of
noncanonical or nonunitary theories for the representation of systems at a fixed value of time
without any possible dynamics. In this case no irreversibility can be consistently treated
due to the need of a time evolution for its very manifestation.

The physical inconsistencies are equally catastrophic, as illustrated by occurrences 1) to 5)
of Theorem 1.3 (see Ref. [29] for details). It is sufficient here to recall that the basic unit
of the three-dimensional  Euclidean space $I = Diag. (1, 1, 1)$ represents in an abstract
dimensionless  form  the assumed  {\it measurement units,} e.g., $I = Diag. (1 cm, 1 cm,
cm)$. The loss of the  measurement units under the time evolution then illustrates the catastrophic
character of the inconsistencies due, e.g.,  to the consequential lack of invariant numerical
predictions.

Similarly, it is easy to prove that the condition of Hermiticity at the
initial time,
$$
(\langle \phi | \times H^\dag )\times | \psi \rangle\equiv
\langle \phi | \times (H\times | \psi \rangle), \; \; H = H^\dag,
\eqno(1.28)
$$
is violated at subsequent
times for theories with nonunitary time evolution when formulated on ${\cal H}$
over $\cal C$. This additional catastrophic inconsistency
(known as {\sl Lopez's lemma} [31-32]), can be expressed by
$$
[\langle \psi |\times W^\dag\times (W\times W^\dag )^{-1}\times
 W\times H\times W^\dag ]\times W | \psi \rangle =
$$
$$
= \langle \psi |\times W^\dag\times [(W\times H\times W^\dag )\times
(W\times W^\dag )^{-1}\times W\times | \psi \rangle] =
$$
$$
=  (\langle \hat\psi\times T\times H'^\dag)\times
| \hat\psi\rangle =
\langle \hat\psi |\times (\hat H\times T\times
| \hat\psi \rangle),
\eqno(1.29a)
$$
$$
 | \hat\psi \rangle = W\!\times\! | \psi \rangle,\; \;
T = (W\!\times\! W^\dag)^{-1} = T^\dag,
\eqno(1.292b)
$$
$$
H'^\dag  =  T^{-1} \!\times\! \hat H \!\times\! T \ne H.
\eqno(1.29c)
$$
As a result, nonunitary theories treated with conventional mathematics do not admit physically
meaningful observables.

Perhaps more insidious is the catastrophic inconsistency caused by the general violation of
causality by theories with nonunitary time evolutions since, the verification of causality laws
by quantum mechanics is deeply linked to the unitarity of its time evolution, as well known.

By no means the above catastrophic inconsistencies solely apply to Santilli's early
formulations of irreversibility. In fact, due to the ''direct universality'' of theories
(1.20), the same inconsistencies apply to a rather vast number of theories, such:

 1) Dissipative nuclear theories [25] represented via an imaginary
potential in non-Hermitean Hamiltonians,
$$
H = H_0 = iV \ne H^\dag
\eqno(1.30)
$$
  lose
all algebras in the brackets of their time evolution (requiring a
bilinear product) in favor of the triple system,
$$
i\times \ dA/dt = A\times H -
H^\dag\times A = [A, H, H^\dag ]
\eqno(1.31)
$$
This causes the loss of nuclear
notions such as ``protons and neutrons'' as conventionally understood,
e.g., because the definition of their spin mandates
the presence of a consistent algebra in the brackets of the time
evolution.

 2) Statistical theories with an external collision term $C$ (see Ref. [38] and literature quoted
therein) and equation of the density
$$
i\ d\rho /dt = \rho\odot H = [\rho , H] + C, \; \;H = H^\dag,
\eqno(1.32)
$$
violate the conditions for the product $\rho\odot H$ to characterize any
algebra, as well as the existence of exponentiation to a finite transform, let
alone violating the conditions of unitarity.

 3) The so-called ``$q$-deformations'' of the Lie product (see, e.g., [39-44] and very
large literature quoted therein)
$$
A\times B - q\times B\times A,
\eqno(1.33)
$$
where $q$ is a non-null scalar, that are a trivial particular case
of Santilli's $(p, q)$-deformations (1.11) introduced in 1967 [7].

 4) The so-called ``$k$-deformations'' [45-48] that are a relativistic
version of the $q$-deforma\-tions, thus also being a particular case of
general structures (1.4.42).

 5) The so-called ``star deformations'' [49] of the
associative product
$$
A \star B = A\times T\times B,
\eqno(1.34)
$$
where  $T$ is fixed, and related generalized Lie product
$$
A \star B - B \star A,
\eqno(1.35)
$$
are manifestly nonunitary and {\sl coincide}
with Santilli's Lie-isotopic algebras introduced in 1978 [11,12].

 6) Deformed creation-annihilation operators theories [50,51].

 7) Nonunitary statistical theories [52].

 8) Irreversible \hfill  black \hfill holes \hfill dynamics \hfill [53] with
 \hfill  Santilli's \hfill Lie-admissible\\
structure (1.20) .

 9) Noncanonical time theories [54-56].

 10) Supersymmetric theories [57] with product
$$
(A, B) =[A, B] + \{ A,B \} =
$$
$$
=  (A\times B - B\times A) +
(A\times B + B\times A),
\eqno(1.36)
$$
are an evident
particular case of Santilli's Lie-admissible product (1.4.46) with
$T = W = I$.

 11) String theories (see ref. [37] and literature quoted therein) generally have a noncanonical
structure due to the inclusion of gravitation with additional catastrophic inconsistencies when
including supersymmetries.

12) The so-called squeezed states theories [58,59] due to their manifest nonunitary
character.

13) All quantum groups (see, e.g., refs. [60-62]) with a nonunitary structure.

 14) Kac-Moody superalgebras [63] are also nonunitary and a particular case
of Santilli's Lie-admissible algebra (1.20) with $T = I$ and $Q$ a phase
factor.

Numerous additional theories are also afflicted by the catastrophic inconsistencies of Theorem 1.3, such as
quantum groups, quantum gravity, and other theories with nonunitary time evolution formulated on
conventional Hilbert spaces over conventional fields the reader can easily identify in the literature.

{\sl  All the above theories have a nonunitary structure formulated via
conventional mathematics and, therefore, are afflicted by the catastrophic mathematical and physical
inconsistencies of Theorem~1.3.}

 Additional generalized theories were attempted via {\bf\sl the relaxation of
the linear character of quantum mechanics} [35]. These theories are
essentially based on eigenvalue equations with the structure
$$
H(t, r, p,| \psi \rangle )\times | \psi \rangle = E\times | \psi \rangle,
\eqno(1.37)
$$
(i.e., $H$ depends on the wavefunction).

 Even though mathematically intriguing and possessing a seemingly
unitary time evolution, these theories also possess rather serious
physical drawbacks, such as:  they violate the superposition principle
necessary for composite systems such as a hadron; they violate the fundamental Mackay
imprimitivity theorem  necessary for the applicability
of Galileo's and Einstein's relativities and possess other drawbacks [18b] so
serious to prevent consistent applications.

 Yet another type of broader theory is {\bf\sl Weinberg's nonlinear theory} [64] with brackets of the type
$$
A \odot B - B \odot A =
$$
$$
=  \frac{\partial A}{\partial \psi}\times \frac{\partial B}{\partial \psi^\dag} -
\frac{\partial B}{\partial \psi}\times \frac{\partial A}{\partial \psi^\dag},
\eqno(1.38)
$$
where the product $A\odot B$  is {\it nonassociative.}

This
theory violates Okubo's No-Quantization Theorem [30], prohibiting the use of nonassociative envelopes
because of catastrophic physical consequences, such as the loss of equivalence between the
Schr\"odinger and Heisenberg representations (the former remains
associative, while the latter becomes nonassociative, thus resulting in
inequivalence).

Weinberg's theory also suffers from the absence of any unit at all, with consequential inability to apply the
theory to measurements, the loss of exponentiation to a finite transform (lack of Poincar\'e-Birkhoff-Witt
theorem), and other inconsistencies studied in Ref. [11].

These inconsistencies are not resolved by the adaptation of Weinberg's theory proposed by Jordan [65] as
readers seriously interested in avoiding the publication of theories known to be inconsistent {\it ab initio}
are encouraged to verify.

 In conclusion, by the late 1970's Santilli had identified classical
and operator generalized theories [11,12] that were proved to be directly universal and include as trivial
particular cases  a plethora of simpler versions by various
other authors.

However, all these theories subsequently resulted in being catastrophically
inconsistent on mathematical
and physical grounds because they are noninvariant under their time evolution when elaborated with
conventional  mathematics, thus
mandating the search for a new mathematics capable of restoring the invariance of quantum
mechanics.

\vskip0.30cm


\noindent {\bf 1.5 Guide to the Research.}
The first need for a serious study in the problem considered is the{\it  identification of a new
mathematics specifically constructed for the invariant treatment of irreversibility at the
classical and operator levels.}

As an example,  classical Lie-admissible equatisons (1.15) and (1.22) are manifestly not
derivable from a potential (because variationally nonselfadjoint [6]), and, consequently, they do
not admit a unique and ambiguous map into an operator form.

As a matter of fact, the lack of a universal
representation of nonconservative forced via a variational principle is a main historical reason for
the inability of the physics of the 20-th century to conduct quantitative studies on the origin of
irreversibility.

As a result, a primary objective of the needed new mathematics is that of achieving a directly
universal representation of all (sufficiently smooth) nonconservative and irreversible systems via a
variational principle. The needed new mathematics will be presented in the next section. Our classical
invariant treatment of irreversibility is presented in Section 3 and its operator counterpart is
presented in Section 4.

The second need for a serious study in the origin of irreversibility is the {\it inclusion of
antimatter ab initio.}

Note that  all the preceding theories were solely intended for the representation of
matter and are inapplicable to antimatter for numerous reasons. For instance,  classical
equations (1.15) and (1.22) permit no differentiation at all between neutral matter and
antimatter stars. Assuming that said equations admit a sort of operator map, the operator image of a
classical representation of antimatter with the sole change of the sign of the charge (as solely
done in the 20-th century) would be given by a ''particle'' with the wrong sign of the charge and
definitely not by a charge conjugated ''antiparticle.''

In any case, one of the historical scientific imbalances of the 20-th
century has been the treatment of matter at all possible levels, from Newtonian mechanics to quantum
field theory, while antimatter was treated at the sole level of second quantization. In turn this
imbalance has prohibited the initiation of the study whether a far away galaxy or quasar is made up of
matter of antimatter (since such a study requires a consistent classical gravitational theory of
antimatter with null total charge, thus without the use of the charge for the characterization of
antimatter).

In view of these and other insufficiencies, the author had to conduct a separate laborious
construction of a new mathematics specifically conceived for the treatment of antiparticle at the
{\it classical} level in such a way to yield operator images equivalent to charge conjugated states.

This new mathematics, today known as {\it Santilli's isodual mathematics,}  is based on the
application of the  {\it isodual map,} here generically expressed by
$$
Q(t, r, \psi, \partial \psi, ...)  \rightarrow  Q^d(t^d, r^d, \psi^d, \partial^d
\psi^d, ...) = - Q^\dag(-t^\dag, -r^\dag, -\psi^\dag, -\partial^\dag (- \psi^\dag), ...)
\eqno(1.39)
$$
applied to the {\it totality} of the mathematical and physical formulations used for matter,
resulting in this way in the novel {\it isodual numbers, isodual fields, isodual spaces, isodual
differential calculus, isodual functional analysis, isodual geometries, Lie-Santilli isodual
theory, isodual symmetries, etc.}.

The classical and operator isodual theory of antimatter cannot possibly be reviewed in this paper to
avoid a prohibitive length. We must, therefore, refer the reader to the existing literature,
such as monographs [18,22] (see monographs [66-72] for independent studies) and vast literature
quoted therein.

The reader should be alerted that  the
restriction of the studies on the origin of irreversibility solely to matter, as done in the 20-th
century, is insufficient and actually insidious because, as now known for grand unifications [22],
certain basic insufficiencies emerge only when the study of antimatter is included ab initio.

The third need for a serious study of irreversibility is the {\it  joint consideration of
open, nonconservative and irreversible as well as closed, conservative and irreversible systems.}

As a
matter of fact, some of the most relevant irreversible systems can be considered as isolated from the
rest of the universe at the classical level (such as the study of the structure of Jupiter) as well as
at the operator level (such as the study of the structure of a star), in which case these systems
verify all ten conventional total conservation laws, while being intrinsically irreversible.

In view of such a need, {\it Santilli proposed his Lie-admissible formulations as a covering, not of
conventional Lie formulation, but as a covering of the broader
irreversible Lie-isotopic formulations} [11,12]. The latter were originally based on the following
operator equations
$$
i\times {dA\over dt} = A \hat \times H - H \hat \times A =
$$
$$
= A\times T(t, r, \psi, \partial \psi,
...)\times H - H\times T(t, r, \psi, \partial \psi,\times A = [A, H]^*,
\eqno(1.40a)
$$
$$
A(t) = W(t)\times A(0)\times W^\dag(t) = (e^{i\times H\times T\times t})\times A(0)\times
(e^{-i\times t\times T\times H}),
\eqno(1.40b)
$$
$$
W\times W^\dag \not = I, \; \; \;
T = T^\dag > 0,\; \; \;  H = H^\dag,
\eqno(1.40c)
$$
with classical counterpart given by Birkhoff's equatsions [6b]
$$
{dA\over dt} = {\partial A\over \partial b^\mu}\times \Omega_{\mu\nu}(t, b)\times {\partial H(b)\over
\partial b^\nu} = [A, H]^*.
\eqno(1.41)
$$

In both cases the brackets of the time evolution $[A, H]^*$ are totally antisymmetric, thus
permitting the conservation law of the energy
$$
i\times {dH\over dt} = [H, H]^* \equiv 0,
\eqno(1.42)
$$
as well as all other conventional conservation laws [18], under a full representation of
irreversibility given by
$$
T(t, r, \psi, \partial \psi, ...) \not =  T(- t, r, \psi, \partial \psi, ...).
\eqno(1.43)
$$
In particular, the generalized brackets $[A, B]^*$ verify the Lie axioms, although they have a
nontrivial generalized structure (e.g., due to the general noncommutativity of H and T) and they
characterize a formulation, today known as {\it Lie-Santilli isotheory} [6b,11,18,22-72] that
includes {\it isoenveloping algebras, Lie-Santilli isoalgebras, Lie-Santilli isogroups,
isorepresentation theory, isotransformation theory, etc.}

We cannot possibly review in this paper the latter formulations and are regrettably forced to refer
the reader to the existing literature (see, e.g., monographs [18] and vast contributions quoted
therein). For the limited scope of this paper we merely indicate that closed irreversible conditions
are generally obtained by restricting the Lie-admissible brackets to be antisymmetric, yet
generalized form. As an illustration, this
condition is reached for the operator case by requiring that the generally different $R$ and $S$
operators coincide, are Hermitean and actually positive definite (for certain topological conditions
required by the underlying isotopology),
$R = S = T = T^\dag > 0$.

In summary, in this paper we shall consider: 1) Conventional, closed and reversible systems of
particles represented with conventional symbols such as $t, r, H,$ etc., conventional associative
product
$\times$ and conventional Lie theory; 2) Closed irreversible systems of particles represented with
 $\hat{}$ over conventional symbols,  and the Lie-Santilli
isotheory; 3) Open irreversible systems of particles moving forward in time represented with the
upper index $^>$, and related Santilli's Lie-admissible theory; 4) Open irreversible systems
of particles moving backward in time represented with the
upper index $^<$, and related Santilli's Lie-admissible theory; and 5) The antiparticle
counterpart of all preceding systems 1-4 represented with the superscript $^d$ denoting isoduality.

Needless to say, the polyhedric
character of problem as well as the complexity of each individual aspect, are such that, despite
the decades of studies by the author,  studies the origin of irreversibility are at their
infancy because so much remains to be done. It is hoped that this paper will stimulate young minds of
any age to contribute to one of the ultimate frontiers of knowledge.

\vskip0.50cm


\noindent {\bf 2. ELEMENTS OF SANTILLI GENOMATHEMATICS AND ITS ISODUAL}

\noindent {\bf 2.1 Genounits, Genoproducts and their Isoduals}
In the author's view, there cannot be a truly new physical theory without a new
mathematics, and there
cannot be a truly new mathematics without new numbers. Therefore, the resolution
of the problem of
invariance for nonunitary theories required laborious efforts in the search of new
numbers capable of
representing irreversibility via their own axiomatic structure.

After failed attempts and a futile search in the mathematical literature,
Santilli proposed in Refs. [11,12] of 1978 the construction of a {\it new
mathematics} specifically conceived for the indicated task, whose number theory eventually
reached mathematical maturity  only in paper~[13] of 1993,
mathematical maturity for the new differential calculus  in memoir [14]
of 1996 and, finally, an invariant formulation of Lie-admissible
formulations only in paper [15] of 1997.

The new Lie-admissible mathematics is today known as {\it Santilli
genomathematics} [66-76], where the prefix ``geno'' suggested in the original
proposal [11,12] is used in the Greek meaning of ``inducting" new
Axioms (as compared to the prefix ``iso'' of the preceding chapter denoting
the preservation of the axioms).

The basic idea is to lift the isounits of the Lie-isotopic theory [18] into a
form that is still nowhere singular, but  {\it non-Hermitean}, thus
implying the existence of {\it two} different generalized units, today
called {\it Santilli genounits} for the description of matter, that are
generally written [13]
$$
\hat I^> = 1 / \hat T^>, \; \; \; ^<{\hat I} = 1 / ^<{\hat T},
\eqno(2.1a)
$$
$$ \; \; \;
\hat I^> \not = ^<{\hat I}, \; \; \;
\hat I^> = (^<\hat I)^\dag,
\eqno(2.1b)
$$
with two additional {\it isodual genounits} for the description of
antimatter~[14]
$$
(\hat I^>)^d = -  {(\hat I^>)}^\dag = - ^<{\hat I} =  -1 / ^<{\hat T}, \; \; \;
(^<{\hat I})^d = - \hat I^> = - 1 / \hat T^>.
\eqno(2.2)
$$

Jointly, all conventional and/or isotopic products $A\hat \times B$ among
generic quantities (numbers, vector fields, operators, etc.) are lifted in
such a form to admit the genounits as the correct left and right units at
all levels, i.e.,
$$
A > B = A\times \hat T^>\times B,\; \; \; A > \hat I^> = \hat I^> > A = A,
\eqno(2.3a)
$$
$$
A < B = A\times  ^<\hat T\times B,\; \; \; A < ^<{\hat I} = ^<{\hat I} < A
= A,
\eqno(2.3b)
$$
$$
A >^d B = A\times {\hat T^>}{}^d \times B,\; \; \; A >^d {\hat I^>}{}^d = {\hat I^>}{}^d
>^d A = A,
\eqno(2.3c)
$$
$$
A <^d B = A\times  ^<\hat T^d\times B,\; \; \; A <^d{} ^<{\hat I}^d ={} ^<{\hat
I}^d{} <^d A = A,
\eqno(2.3d)
$$
for all elements $A$, $B$ of the set considered.

As we shall see in Section 3, the above basic assumptions permit the
representation of irreversibility with the most primitive possible
quantities, the basic units and related products.

In particular,   genounits
permit an invariant representation of the external forces in Lagrange's and Hamilton's
equations. As such, they are generally dependent on time, coordinates, momenta.,
wavefunctions and any other needed variable, e.g., $\hat I^> = \hat I^>(t^>, r^>, p^>,
\psi^>, \ldots)$.

The assumption of all {\it ordered product to the right} $>$ permits the representation of
matter systems moving forward in time, the assumption of all {\it
ordered products to the left}
$<$ can represent matter systems moving backward in time, with corresponding antimatter systems
represented by the respective isodual ordered products $>^d = - >^\dag$ and $<^d = - <>^\dag$.
Irreversibility is represented {\it ab initio} by the inequality $A>B
\not = A<B$ for matter and $>^d \not = <^d$ for antimatter.

Note that the simpler isotopic cases are given by $\hat I^> = ^<\hat I = \hat I = \hat I^\dag > 0$ for
matter and ${\hat{I}}^{>d} = ^<\hat I^d = \hat I^d = {\hat {I}}^{d\dag} < 0$ for antimatter.

In conclusion, the reader should be aware that genomathematics consists of {\it four} branches, the
{\it forward and backward genomathematics for matter and their isoduals for antimatter,} each paid
being interconnected by time reversal, the two pairs being interconnected by isodual map (1.39) that,
as it is well known [22], is equivalent to charge conjugation.


\vskip0.30cm

\noindent {\bf 2.2. Genonumbers,  Genofunctional Analysis and their Isoduals}
Genomathematics began to reach maturity with the discovery made,
apparently for the first time in paper [13] of 1993, that {\it the axioms of
a field still hold under the ordering of all products to the right or,
independently, to the left.}

This unexpected property permitted the formulation of {\it new numbers,}
that can be best introduced as a generalization of the isonumbers [18], although they can also be
independently presented as follows:
\vskip0.30cm

{\it DEFINITION 2.1 [13]:  Let $F\; =\; F(a,+,\times )$ be a field of characteristic
zero as per Definitions 2.2.1 and 3.2.1.  Santilli's  forward genofields are
rings $\hat F^> \;=\; \hat F(\hat a^>,\hat +^>,\hat {\times}^>)$ with: elements
$$
\hat a^> = a\times \hat I^>,
\eqno(2.4)
$$
where $a\in F$, $\hat I^> = 1/\hat T^>$ is a
non singular  non-Hermitean quantity (number, matrix or operator) generally outside $F$ and
$\times$ is the ordinary product of
$F$; the genosum $\hat +^>$ coincides with the ordinary sum $+$,
$$
\hat a^> \hat +^>\hat b^> \equiv \hat a^> + \hat b^>,\;\; \forall \hat a^>,\; \hat b^> \in \hat F^>,
\eqno(2.5)
$$
consequently, the additive forward genounit $\hat 0^> \in \hat F$ coincides with the ordinary
$0\in F$; and the forward genoproduct $>$ is such that $\hat I^>$ is the right and left
isounit of $\hat F^>$,
$$
\hat I^>\hat \times \hat a^> = \hat a^> >\hat I^> \equiv \hat a^>,\; \;
\forall \hat a^> \in \hat F^>.
\eqno(2.6)
$$

Santilli's forward genofields verify the following properties:

1) For each element $\hat a^> \in \hat F^>$ there is an element ${\hat a^>}{}^{-\hat 1^>}$, called
{\it forward genoinverse,} for which
$$
\hat a^> > {\hat a^>}{}^{-\hat I^>} = \hat I^>,\;\; \forall \hat a^> \in \hat F^>;
\eqno(2.7)
$$

 2) The genosum is commutative
$$
\hat a^> \hat +^> \hat b^> = \hat b^> \hat +^> \hat a^>,
\eqno(2.8)
$$
and associative
$$
(\hat a^> \hat +^> \hat b^>) +^> \hat c^> = \hat a^> \hat +^> (\hat b^> \hat +^> \hat c^>),
\; \; \forall \hat a, \hat b, \hat c \in \hat F;
\eqno(2.9)
$$

3) The forward genoproduct is associative
$$
\hat a^> > (\hat b^> > \hat c^>) = (\hat a^> > \hat b^>) > \hat c^>, \; \; \; \forall \hat
a^>, \hat b^>, \hat c^> \in \hat F^>;
\eqno(2.10)
$$
but not necessarily commutative
$$
\hat a^> > \hat b^> \ne \hat b^> >  \hat a^>,
\eqno(2.11)
$$

4) The set $\hat F^>$ is closed under the genosum,
$$
\hat a^> \hat +^> \hat b^> = \hat c^> \in \hat F^>,
\eqno(2.12)
$$
the forward genoproduct,
$$
\hat a^> > \hat b^> = \hat c^> \in \hat F^>,
\eqno(2.13)
$$
and right and left genodistributive compositions,
$$
\hat a^> > (\hat b^> \hat +^> \hat c^>) = \hat d^> \in \hat F^>,
\eqno(2.14a)
$$
$$
(\hat a^> \hat +^> \hat b^>) >
 \hat c^> = \hat d^> \in \hat F^>\; \; \;  \forall \hat a^>, \hat b^>, \hat c^>, \hat d^>
\in \hat F^>;
\eqno(2.14b)
$$

5) The set $\hat F^>$ verifies the right and left genodistributive law
$$
\hat a^> > (\hat b^> \hat +^> \hat c^>) = (\hat
a^> \hat +^> \hat b^>) > \hat c^> = \hat d^>,  \; \;  \forall \hat a^>, \hat b^>, \hat c^>,
\in \hat F^>.
\eqno(2.15)
$$

In this way we have the forward genoreal numbers $\hat R^>$, the forward genocomplex numbers $\hat
C^>$ and the forward genoquaternionic numbers $\hat QC^>$ while the forward genooctonions $\hat O^>$
can indeed be formulated but they do not constitute genofields [14].

The backward genofields and
the isodual forward and backward geno\-fields are defined accordingly. Santilli's genofields are
called of the  first (second) kind when the genounit is (is not) an element of
F. }

\vskip 0.30cm

The basic axiom-preserving character of genofields is illustrated by the
following:
\vskip0.50cm

{\it LEMMA 2.1 [13]: Genofields of first and second kind are fields (namely,
they verify all axioms of a field).}

\vskip 0.50cm

Note that the conventional product ``2
multiplied by 3'' is not necessarily equal to 6 because, for isodual numbers with unit $-1$ it is given by $-6$ [13].

The same product ``2 multiplied
by~3'' is not necessarily equal to +6 or $-6$ because, for the case of isonumbers, it can also
be equal to an arbitrary number, or a matrix or an
integrodifferential operator depending on the assumed isounit [13].

In this section we point out that ``2 multiplied by 3'' can be
ordered to the right or to the left, and the result is not only
arbitrary, but yielding different numerical results for different
orderings,  $2 > 3 \not = 2 < 3$, all this by continuing to verify
the axioms of a field per each order [13].

Once the forward and backward genofields have been identified, the
various branches of genomathematics can be constructed via simple
compatibility arguments.

For specific applications to irreversible processes there is first the need to construct the {\it
genofunctional analysis,} studied in Refs. [6,18] that we cannot review here for brevity. the
reader is however warned that any elaboration of irreversible processes via Lie-admissible
formulations based on conventional or isotopic functional analysis leads to catastrophic
inconsistencies because it would be the same as elaborating quantum mechanical calculations with
genomathematics.

As an illustration, Theorems 1.5.1 and 1.5.2 of catastrophic inconsistencies are activated unless
one uses the ordinary differential calculus is lifted, for ordinary motion in time of matter, into
the following {\it forward genodifferentials and genoderivatives}
$$
\hat d^>x = \hat T^>_x\times dx, \ \ {\hat \partial^>\over\hat
\partial^> x} = \hat I^>_x\times
{\partial\over \partial x}, \ {\rm etc.}
\eqno(2.16)
$$
with corresponding backward and isodual expressions here ignored,

Similarly, all conventional functions and isofunctions, such as isosinus, isocosinus, isolog,
etc.,  have to be lifted in the genoform
$$
\hat f^>(x^>) = f(\hat x^>)\times \hat I^>,
\eqno(2.17)
$$
where one should note the necessity of the multiplication by the genounit as a condition for the
result to be in $\hat R^>$, $\hat C^>$, or $\hat O^>$.
\vskip0.30cm


\noindent {\bf 2.3. Genogeometries and Their Isoduals}
Particularly intriguing are the
{\it genogeometries} [16] (see also monographs [18] for
detailed treatments). They are best  characterized
by a simple genotopy of the isogeometries, although they can be independenntly defined.

{\sloppy As an illustration, the {\it Minkowski-Santilli forward genospace} $\hat
M^>(\hat x^>$, $\hat \eta^>,\hat R^>)$ over the  genoreal $\hat R^>$ is
characterized by the following spacetime, {\it genocoordinates, genometric and
genoinvariant}
$$
\hat x^> = x\hat I^> = \{x^\mu\}\times \hat I^>, \; \; \; \hat \eta^> = \hat
T^>\times \eta, \; \; \; \eta = Diag. (1, 1, 1, -1),
\eqno(4.18a)
$$
$$
 {{\hat{x}}^>}{}^{2^>} = {{\hat{x}}^>}{}^{\mu} \hat \times^> \hat \eta^{>}_{\mu\nu}\hat \times ^>
 {{\hat{x}}^>}{}^{\nu} = (x^\mu\times \hat \eta^{>}_{\mu\nu}\times x^\nu)\times \hat
I^>,
\eqno(2.18b)
$$
where the first expression of the genoinvariant is on genospaces while the
second is its projection in our spacetime.

}Note that the minkowski-Santilli genospace has, in general, an explicit
dependence on spacetime coordinates. Consequently, it is equipped with the
entire formalism of the conventional Riemannian spaces covariant
derivative, Christoffel's symbols, Bianchi identity, etc. only lifted
from the isotopic form of the preceding chapter into the genotopic form.

A most important feature is that {\it genospaces permit, apparently for the first time in
scientific history, the  representation of irreversibility directly via the basic genometric.} This
is due to the fact that genometrics are nonsymmetric by conception, e.g.,
$$
\hat \eta^>_{\mu\nu} \; \; \; \not = \; \; \; \hat \eta^>_{\nu\mu}.
\eqno(2.19)
$$

Consequently, {\it genotopies permit the lifting of conventional
symmetric metrics into nonsymmetric forms,}
$$
\eta^{Minkow.}_{Symm} \; \; \; \rightarrow \; \; \; \hat \eta^>{}^{Minkow.-Sant.}_{NonSymm}
\eqno(2.20)
$$

Remarkably, {\it nonsymmetric metrics bare indeed permitted by the axioms of
conventional spaces} as illustrated by the invariance
$$
(x^\mu\times \eta_{\mu\nu}\times x^\nu)\times I \equiv
[x^\mu\times (\hat T^>\times \eta_{\mu\nu})\times x^\nu]\times T^>{}^{-1}
\equiv
$$
$$
\equiv (x^\mu\times\hat
\eta^>_{\mu\nu}\times x^\nu)\times \hat I^>,
\eqno(2.21)
$$
where $\hat T^>$ is assumed in this simple illustration to be a complex
number.

Interested readers can then work out backward genogeometries and the isodual forward and backward
genogeometries with their underlying genofunctional analysis.

This basic geometric feature was not discovered until
recently because hidden where nobody looked for, in the basic unit.
However, this basic geometric advance in the representation of
irreversibility required the prior discovery of basically new numbers,
Santilli's genonumbers with nonsymmetric unit and ordered multiplication [14].

\vskip0.30cm


\noindent {\bf 2.4. Santilli Lie-Admissible heory and its Isodual}
Particularly important for irreversibility is the lifting of Lie's theory
and Lie-Santilli's isotheories permitted by genomathematics, first
identified by  Ref. [11] of 1978 (and then  studied in various works, e.g., [6,18-22])
via the following genotopies:

(1) The {\it forward and backward universal enveloping
genoassociative algebra} $\hat \xi^>,\; ^<\hat {\xi}$, with
infinite-dimensional basis characterizing the {\it
Poin\-car\'e-Birkhoff-Witt-Santilli genotheorem }
$$
\hat \xi^>: \hat I^>,\; \hat X_i,\; \hat X_i > \hat X_j,\; \hat X_i > \hat
X_j > \hat X_k,\; \ldots,\; i\leq j \leq k,
\eqno(2.22a)
$$
$$
^<\hat \xi: \hat I,\; ^<\hat X_i,\; \hat X_i < \hat X_j,\; \hat X_i < \hat
X_j < \hat X_k,\; \ldots,\; i\leq j \leq k;
\eqno(2.22b)
$$
where the ``hat'' on the generators denotes their formulation on
geno\-spaces over genofields and their Hermiticity implies that
$\hat X^> = ^<\hat X = \hat X$;

(2) The {\it Lie-Santilli genoalgebras} characterized by the
universal, jointly Lie- and Jordan-admissible brackets,
$$
^<\hat L^>: (\hat X_i\hat , \hat X_j) = \hat X_i < \hat X_j - \hat
X_j
> \hat X_i =
 C_{ij}^k\times \hat X_k,
\eqno(2.23)
$$
here formulated formulated in an invariant form (see below);

(3) The {\it Lie-Santilli genotransformation groups}
$$
^<\hat G^>: \hat A(\hat w) = (\hat e^{\hat i\hat {\times} \hat
X\hat {\times}\hat w})^>
> \hat A(\hat 0) < ^<(\hat e^{-\hat i\hat {\times}\hat w\hat {\times}
\hat X}) =
$$
$$
= (e^{i\times \hat X\times \hat T^>\times  w})\times A(0)\times
(e^{-i\times w\times ^<\hat T\times \hat X}),
\eqno(2.24)
$$
where $\hat w^> \in \hat R^>$ are the {\it genoparameters;} the {\it
genorepresentation theory,}  etc.

\vskip0.30cm


\noindent {\bf 2.5. Genosymmetries and Nonconservation Laws}
The  implications of the Santilli Lie-admissible theory are
significant mathematically and physically. On mathematical grounds,
the Lie-Santilli genoalgebras are ``directly universal'' and include
as particular cases all known algebras, such as Lie, Jordan,
Flexible algebras, power associative algebras, quantum, algebras,
supersymmetric algebras, Kac-Moody algebras, etc. (Section 1.5).

Moreover,  when computed on the {\it genobimodule}
$$
^<\hat B^> = ^<\hat {\xi}\times \hat {\xi}^>,
\eqno(2.25)
$$
 {\it Lie-admissible algebras verify all Lie axioms,} while deviations
from Lie algebras emerge only in their {\it projection} on the conventional
bimodule
$$
^<B^> = ^<\xi\times \xi^>,
\eqno(2.26)
$$
 of Lie's theory (see Ref. [17] for the initiation of the genorepresentation theory of
Lie-admissible algebras on bimodules).

This is due to the fact that the computation of the left action $A <
B = A\times ^<\hat T\times B$ on $^<\hat {\xi}$ (that is, with
respect to the genounit $^<\hat I = 1/^<\hat T$) yields the save
value as the computation of the conventional product $A\times B$ on
$^<\xi$ (that is, with respect to the trivial unit $I$), and the
same occurs for the value of $A > B$ on $\hat {\xi}^>$.

The above occurrences explain the reason the structure constant and
the product in the r.h.s. of Eq. (4.2.23) are those of a conventional
Lie algebra.

In this way, thanks to genomathematics, {\it Lie algebras acquire a
towering significance in view of the possibility of reducing all
possible irreversible systems to primitive Lie axioms.}

The physical implications of the Lie-Santilli genotheory are equally
far reaching. In fact, Noether's theorem on the reduction of
reversible conservation laws to primitive Lie symmetries can be lifted
to the {\it reduction, this time, of irreversible nonconservation
laws to primitive Lie-Santilli genosymmetries.}

As a matter of fact, this reduction was the very first motivation
for the construction of the genotheory in memoir [12] (see also
monographs [6,18,19,20]). The reader can then foresee similar liftings of
all remaining physical aspects treated via Lie algebras.

The construction of the isodual Lie-Santilli genotheory is an
instructive exercise for readers interested in learning the new
methods.

\vskip0.50cm

\noindent {\bf 3.  LIE-ADMISSIBLE CLASSICAL MECHANICS FOR MATTER AND ITS ISODUAL FOR ANTIMATTER}

\noindent {\bf 3.1. Fundamental Ordering Assumption on Irreversibility}
Another reason for the inability during the 20-th century for in depth studies of irreversibility is
the general belief that motion in time has only two directions, forward and backward (Eddington
historical time arrows). In reality, motion in time admits {\it four} different forms, all essential
for serious studies in irreversibility, given by: 1) {\it motion forward to future time}
characterized by the forward genotime $\hat t^>$; 2) {\it motion backward to past time}
characterized by the backward genotime $^<\hat t$; 3) {\it motion backward from future time}
characterized by the isodual forward genotime $\hat{t}^{>d}$; and 4) {\it motion forward from past time}
characterized by the isodual backward genotime $^<\hat t^d$.

It is at this point where the {\it necessity} of both time reversal and isoduality appears in its
full light. In fact, time reversal is only applicable to matter and, being represented with Hermitean
conjugation, permits the transition from motion forward to motion backward in time, $\hat
t^>\rightarrow ^<\hat t = (\hat t^>)^\dag$. If used alone, time reversal cannot identify all four
directions of motions. The {\it only} additional conjugation known to this author that is applicable
at alllevels of study and is equivalent to charge conjugation, is isoduality [22].

The additional discovery of two complementary  orderings of the
product and related units, with corresponding isoduals versions, individually preserving the abstract
axioms of a field has truly fundamental implications for irreversibility, since it permits the
axiomatically consistent and invariant representation of
irreversibility via the most ultimate and primitive axioms, those on
the product and related unit. We, therefore, have the following:
\vskip0.50cm

{\it FUNDAMENTAL ORDERING ASSUMPTION ON IRREVER\-SI\-BI\-LI\-TY} [15]: {\it
Dynamical equations for motion forward in time of matter (antimatter) systems are characterized by
genoproducts to the right and related genounits (their isoduals), while dynamical
equations for the motion backward in time of matter (antimatter) are characterized by
genoproducts to the left and related genounits (their isoduals) under the condition
that said genoproducts and genounits are interconnected by time
reversal expressible for generic quantities $A$, $B$ with the relation,}
$$
(A > B)^\dag = (A > \hat T^>\times B)^\dag = B^\dag\times (\hat
T^>)^\dag\times A^\dag,
\eqno(3.1)
$$
{\it namely,}
$$
\hat T^> = (^<\hat T)^\dag
\eqno(3.2)
$$
{\it thus recovering the fundamental complementary conditions
(4.1.17) or (4.2.2).}

Unless otherwise specified, from now on physical and chemical expression for irreversible
processes will have no meaning without the selection of one of the indicated two possible
orderings.

\vskip0.30cm


\noindent {\bf 3.2. Geno-Newtonian Equations and Their Isoduals}
Recall that, for the case of isotopies, the basic Newtonian systems
are given by those admitting nonconservative internal forces
restricted by certain constraints to verify total conservation laws
called {\it closed non-Hamiltonian systems} [6b,18].

For the case of the genotopies under consideration here, the basic
Newtonian systems are the conventional nonconservative systems
without subsidiary constraints, known as{\it open non-Hamiltonian systems},
with generic expression (1.3), in which case irreversibility is entirely
characterized by nonselfadjoint
forces, since all conservative forces are
reversible.

As it is well known, the above equations are not derivable from any
variational principle in the fixed frame of the observer [6], and
this is the reason all conventional attempts for consistently
quantizing nonconservative forces  have failed for about one
century. In turn, the lack of  achievement of a consistent operator
counterpart of nonconservative forces lead to the belief
that they are ÒillusoryÓ because they ÒdisappearÓ at the particle level.

The studies presented in this paper have achieved the first and only physically
consistent operator formulation of nonconservative forces known to
the author. This goal was achieved by rewriting
Newton's equations (1.3) into an identical form derivable from a variational principle. Still in
turn, the latter objective was solely permitted by the novel
genomathematics.

It is appropriate to recall that Newton was forced to discover new
mathematics, the differential calculus, prior to being able to
formulated his celebrated equations. Therefore,  readers should not be surprised at the need
for the new genodifferential calculus  as a condition to represent all nonconservative Newton's
systems from a variational principle.

Recall also from Section 3,1 that, contrary to popular beliefs,
there exist {\it four} inequivalent directions of time. Consequently, time reversal alone cannot represent all these
possible motions, and isoduality results to be the only known
additional conjugation that, when combined with time reversal, can
represent all possible time evolutions of both matter and
antimatter.

The above setting implies the existence of four different new
mechanics first formulated by Santilli in memoir [14] of 1996, and
today known as {\it Newton-Santilli genomechanics,} namely:

A) {\it Forward genomechanics} for the representation of forward motion of matter systems;

B)  {\it Backward genomechanics} for the representation of the time
reversal image of matter systems;

C)  {\it Isodual backward genomechanics} for the representation of
motion backward in time of antimatter systems, and

D)  {\it Isodual forward genomechanics} for the representation of
time reversal antimatter systems.

These new mechanics are characterized by:

1) Four different times,  {\it forward and backward genotimes for
matter systems and the backward and forward isodual genotimes for
antimatter systems}
$$
\hat t^> = t\times \hat I^>_t, \; \; \; -\hat t^>, \; \; \;
\hat t^>{}^d,\; \; \; - \hat t^>{}^d,
\eqno(3.3)
$$
with (nowhere singular and non-Hermitean) {\it forward and backward
time genounits and their isoduals} (Note that, to verify the
condition of non-Hermiticity, the time genounits can be {\it complex
valued.}),
$$
\hat I^>_t = 1/\hat T^>_t, \; \; \; -\hat
I^>_t, \; \; \; \hat I^>{}^d_t, \; \; \; - \hat I^>{}^d_t;
\eqno(3.4)
$$

2) The {\it forward and backward genocoordinates and their isoduals}
$$
\hat x^> = x\times \hat I^>_x,\;\;\; - \hat x^>, \; \; \; \hat x^>{}^d, \;
\; \; - \hat x^>{}^d,
\eqno(3.5)
$$
 with (nowhere singular non-Hermitean) {\it
coordinate genounit}
$$
\hat I^>_x = 1/\hat T^>_x, \; \; \; - \hat
I^>_x, \; \; \; \hat I^>{}^d_x, \; \; \; - \hat I^>{}^d_x,
\eqno(3.6)
$$
 with {\it forward and backward coordinate genospace and their
isoduals} $\hat S^>_x$, etc., and related {\it forward coordinate
genofield and their isoduals}  $\hat R^>_x$, etc.;

3) The {\it forward and backward genospeeds and their isoduals}
$$
\hat v^> = \hat d^>\hat x^>/\hat d^>\hat t^>, \; \; \; - \hat v^>,
\; \; \; \hat v^>{}^d, \; \; \; - \hat v^>{}^d,
\eqno(3.7)
$$
 with (nowhere singular and non-Hermitean) {\it
speed genounit}
$$
\hat I^>_v = 1/\hat T^>_v,\; \; \; - \hat
I^>_v, \; \; \; \hat I^>{}^d_v, \; \; \; - \hat I^>{}^d_v,
\eqno(3.8)
$$
 with related
{\it forward speed backward genospaces and their isoduals}  $\hat
S^>_v$, etc.,  over {\it forward and backward speed genofields}
$\hat R^>_v$, etc.;

The above formalism then leads to the {\it forward genospace for matter systems}
$$
\hat S^>_{tot} = \hat S^>_t\times \hat S^>_x\times \hat S^>_v,
\eqno(3.9)
$$
defined over the {it forward genofield}
$$
\hat R^>_{tot} = \hat R^>_t\times \hat R^>_x\times \hat R^>_v,
\eqno(3.10)
$$
with  {\it total forward genounit}
$$
\hat I^>_{tot} = \hat I^>_t\times \hat I^>_x\times \hat I^>_v,
\eqno(3.11)
$$
and corresponding expressions for the remaining three spaces
obtained via time reversal and isoduality.

The basic equations are given by:

I) The {\it forward Newton-Santilli genoequations for matter
systems} [14], formulated  via the genodifferential calculus,
$$
\hat m^>_a > {\hat d^> \hat v^>_{ka}\over \hat d^> \hat
t^>} = -{\hat \partial^> \hat V^>\over \hat \partial^> \hat
x^{>k}_a};
\eqno(3.12)
$$

II) The {\it backward genoequations  for matter systems} that are
characterized by time reversal of the preceding ones;

III) the {\it backward isodual  genoequations for antimatter
systems} that are characterized by the isodual map of the backward
genoequations,
$$
^<{\hat m}^d_a < \frac{{^<{\hat d}^d} {^<{\hat v}^d_{ka}}}
{{^<{\hat d}^d} {^<{\hat t}^d}} = -\frac{{^<\hat \partial^d}
{^<{\hat V}^d}} {{^<\hat \partial^d} {^<{\hat x}^{dk}_a}};
\eqno(3.13)
$$

IV) the {\it forward isodual genoequations for antimatter systems}
characterized by time reversal of the preceding isodual equations.

Newton-Santilli genoequations (4.3.12) are ``directly universal'' for the representation of all
possible (well behaved) Eqs. (1.3) in the frame of the observer because they admit a multiple
infinity of solution for any given nonselfadjoint force.

A simple representation occurs under the conditions assumed for simplicity,
$$
N = \hat I^>_t = \hat I^>_v = 1,
\eqno(3.14)
$$
for which Eqs. (3.12) can be explicitly written
$$
\hat m^> > {\hat d^>\hat v^>\over \hat d^> t}
= m \times{d\hat v^>\over dt} =
$$
$$
= m\times {d\over dt}{d(x\times \hat I^>_x)\over dt} =
m\times {dv\over dt}\times \hat I^>_x + m\times x\times {d\hat I^>\over dt} = \hat
I^>_x\times {\partial V\over
\partial x},
\eqno(3.15)
$$
from which we obtain the genorepresentation
$$
F^{NSA} = - m\times x\times  {1\over \hat I^>_x}\times {d\hat I^>_x\over dt},
\eqno(3.16)
$$
that always admit solutions here left to the interested reader since in the next section we shall
show a much simpler, universal, {\it algebraic} solution.

As one can see,  in Newton's equations
the nonpotential forces are part of the applied force, while in the
Newton-Santilli genoequations nonpotential forces are represented by
the  genounits, or, equivalently, by the  genodifferential
calculus, in a way essentially similar to the case of isotopies.

The main difference between iso- and geno-equations is that isounits
are Hermitean, thus implying the equivalence of forward and backward
motions, while genounits are non-Hermitean, thus implying
irreversibility.

Note also that the topology underlying Newton's equations is the
conventional, Euclidean, local-differential topology which, as such,
can only represent point particles.

By contrast, the topology underlying the Newton-Santilli
genoequations is given by a genotopy of the isotopology studied in
the preceding chapter, thus permitting for the representation of extended,
nonspherical and deformable particles via forward genounits, e.g.,
of the type
$$
\hat I^> = Diag.  (n_1^2, n_2^2, n_3^2, n^2_4)\times \Gamma^>(t, r,
v, \ldots),
\eqno(3.17)
$$
where $n_k^2,\; k = 1, 2, 3$ represents the semiaxes of an ellipsoid,
$n^2_4$ represents the density of the medium in which motion occurs
(with more general nondiagonal realizations here omitted for
simplicity), and $\Gamma^>$ constitutes a nonsymmetric matrix
representing  nonselfadjoint forces, namely, the
contact interactions among extended constituents occurring for the
motion forward in time.
\vskip0.30cm


\noindent {\bf 3.3. Lie-Admissible Classical Genomechanics and its
Isodual}
In this section we show that, once rewritten in their identical
genoform (4.3.12), Newton's equations for nonconservative systems
are indeed derivable from a variational principle, with  analytic
equations possessing a Lie-admissible structure and Hamilton-Jacobi
equations suitable for the first know consistent and unique operator
map studied in the next section.

The most effective setting to introduce real-valued  non-symmetric
genounits is in the $6N$-dimensional {\it forward genospace
(genocotangent bundle)}  with local genocoordinates and their
conjugates
$$
\hat a^{>\mu} = a^{\rho}\times {\hat I}^{>\mu}_{1 {\kern 2pt}\rho},\;\;
(\hat a^{>\mu}) =
\left (
\begin{array} {ccc}
 \hat x^{>k}_\alpha \\
\hat p^>_{k\alpha}
\end{array} \right )
\eqno(3.18)
$$
and
$$
\hat R^>_\mu = R_\rho\times \hat I^{>\rho}_{2{\kern 2pt}\mu}, \
(\hat R^>_\mu) =  (\hat p_{k\alpha}, \hat 0),
\eqno(3.19a)
$$
$$
\hat I_1^> = 1/\hat T_1^> = (\hat I_2^>)^T = (1/\hat T_2^>)^T,
\eqno(3.19b)
$$
$$
k = 1, 2, 3; \ \alpha =  1, 2, \ldots, N; \ \  \mu, \ \rho = 1, 2,
\ldots6N,
$$
where the superscript $T$ stands for transposed, and nowhere
singular, real-valued and
non-symmetric genometric and related invariant\\
$$
\hat \delta^> = \hat T^>_{1{\kern 2pt}
6N\times 6N} {\kern 1pt} \delta_{6N\times 6N} \times \delta_{6N\times 6N} ,
\eqno(3.20a)\\
$$
$$
\hat a^{>\mu} > \hat R^>_\mu = \hat a^{>\rho}\times \hat
T^{>\beta}_{1{\kern 2pt}\rho}\times \hat R^>_\beta = a^\rho\times
\hat I^{>\beta}_{2{\kern 2pt}\rho}\times R_\beta.
\eqno(3.20b)\\
$$

In this case we have the following {\it genoaction principle}  [14]
$$
\hat \delta^> \hat {\cal A}^> = \hat \delta^> \hat {\int}^>[\hat
R^>_\mu >_a \hat d^> \hat a^>{}^\mu - \hat H^> >_t \hat d^> \hat t^>] =
$$
$$
= \delta {\int}[R_\mu\times \hat T^{>{\kern 1pt}\mu}_{1{\kern
2pt}\nu}(t, x, p, \ldots)\times d (a^\beta\times \hat I^{>\nu}_{1{\kern
2pt}\beta}) -
 H\times dt] = 0,
\eqno(3.21)
$$
where the second expression is the projection on conventional spaces
over conventional fields and we have assumed for simplicity that the
time genounit is 1.

It is easy to prove that the above genoprinciple characterizes the
following {\it forward Hamilton-Santilli genoequations,} (originally
proposed in Ref. [11] of 1978 with conventional mathematics and in
Ref. [14] of 1996 with genomathematics (see also Refs.
[18,19,20])
$$
\hat \omega_{\mu\nu}^> > {{\hat d^>}\hat a^{\nu {\kern 1pt}>}
\over\hat d^>\hat t^>} - {\hat \partial^>\hat H^>(\hat a^>)\over\hat
\partial^>\hat a^{\mu {\kern 1pt}>}} =
$$
$$
= \left ( \begin{array} {ccc}
 0 & -1 \\
 1 & 0\\
\end{array} \right ) \times
 \left ( \begin{array} {ccc}
 dr/dt \\
 dp/ dt
\end{array} \right ) -
\left ( \begin{array} {ccc}
 1 & K \\
 0 & 1\\
\end{array} \right ) \times
\left ( \begin{array} {ccc}
 \partial H/\partial r \\
\partial H/\partial p\\
\end{array} \right ) = 0,
\eqno(3.22a)
$$
$$
\hat \omega^> = \Big({\hat \partial^> R^>_\nu \over \hat \partial^>
\hat a^{\mu {\kern 1pt}>}} - {\hat \partial^>\hat R^>_\mu\over \hat
\partial^> \hat a^{\nu {\kern 1pt}>}}\Big)\times \hat I^> = \left(
\begin{array} {ccc}
 0 & -1 \\
 1 & 0\\
\end{array} \right) \times \hat I^>,
\eqno(3.22b)
$$
$$
K = F^{NSA}/(\partial H/\partial p), \eqno(4.3.22c)
$$
where one should note the ``direct universality'' of the simple algebraic solution (3.22c).

The time evolution of a quantity $\hat A^>(\hat a^>)$ on the forward
geno-phase-space can be written in terms of the following brackets
$$
{\hat d^>\hat A^>\over \hat d^>t^>} = (\hat A^>, \hat H^>) = {\hat
\partial^> \hat A^>\over \hat
\partial^>
\hat a^{>\mu}} > \hat \omega^{\mu\nu}{}^> > {\hat \partial^> \hat
H^>\over \hat \partial \hat a^{>\nu}} =
$$
$$
= {\partial \hat A^>\over \partial \hat a^{>\mu}}\times
S^{\>\mu\nu}\times {\partial \hat H^>\over \partial \hat a^{>\nu}} =
$$
$$
= \Big({\partial \hat A^>\over \partial \hat r^{>k}_\alpha}\times
{\partial \hat H^>\over
\partial
\hat p^>_{ka}} - {\partial \hat A^>\over \partial \hat
p^>_{ka}}\times {\partial \hat H^>\over \partial \hat
r^{>k}_a}\Big) + {\partial \hat A^>\over \partial \hat
p^>_{ka}}\times F^{NSA}_{ka},
\eqno(3.23a)
$$
$$
S^{>\mu\nu} = \omega^{\mu\rho}\times \hat I^{2{\kern 1pt \mu}}_\rho,
\omega^{\mu\nu} = (||\omega_{\alpha\beta}||^{-1})^{\mu\nu},
\eqno(3.23b)
$$
where $\omega^{\mu\nu}$ is the conventional Lie tensor and,
consequently, $S^{\mu\nu}$ is Lie-admissible in the sense of Albert
[7].

As one can see, the important consequence of genomathematics and its
genodifferential calculus is that of turning the triple system $(A,
H, F^{NSA})$ of Eqs. (1.5) in the bilinear form $(A\hat , B)$,
thus characterizing a consistent algebra in the brackets of the time
evolution.

This is the central purpose for which genomathematics was built
(note that the multiplicative factors represented by $K$ are fixed
for each given system). The invariance of such a formulation will be
proved shortly.

It is an instructive exercise for interested readers to prove that
the brackets $(A\hat {,} B)$ are Lie-admissible, although not
Jordan-admissible.

It is easy to verify that the above identical reformulation of
Hamilton's historical time evolution correctly recovers the {\it
time rate of variations of physical quantities} in general, and that
of the energy in particular,
$$
{dA^>\over dt} = (A^>, H^>) =  [\hat A^>, \hat H^>] + {\partial \hat
A^>\over
\partial \hat p^{>}_{k\alpha}}\times F^{NSA}_{k\alpha},
\eqno(3.24a)
$$
$$
{dH\over dt} = [\hat H^>, \hat H^>] + {\partial \hat H^>\over
\partial \hat p^{>}_{k\alpha}}\times F^{NSA}_{ka} =
v^k_{\alpha}\times F^{NSA}_{ka}.
\eqno(3.24b)
$$

It is easy to show that genoaction principle (4.3.21) characterizes
the following {\it Hamilton-Jacobi-Santilli genoequations} [14]
$$
{\hat \partial^> {\cal A}^>\over \hat \partial^> \hat t^>} + \hat
H^> = 0,
\eqno(3.25a)
$$
$$
\Bigl({\hat \partial^> {\cal A}^>\over \hat \partial^> \hat a^{>\mu}}\Bigr) =
\Bigl({\hat \partial^> {\cal A}^>\over \hat \partial^> x^{>{\kern
1pt}k}_a}, {\hat \partial^> {\cal A}^>\over \hat \partial^>
p^>_{ka}}\Bigr) = (\hat R^>_\mu) = (\hat p^>_{ka}, \hat 0),
\eqno(3.25b)
$$
which confirm the property (crucial for genoquantization as shown
below) that the genoaction is indeed independent of the linear
momentum.

Note the {\it direct universality} of the Lie-admissible equations
for the representation of all infinitely possible Newton equations (1.3)
(universality) directly in the fixed frame of the experimenter
(direct universality).

Note also that, {\it at the abstract, realization-free level,
Hamilton-Santilli genoequations  coincide} with Hamilton's equations
without external terms, yet represent those with external terms.

The latter are reformulated via genomathematics as the only known
way to achieve invariance and derivability from a variational
principle while admitting a consistent algebra in the brackets of
the time evolution~[38].

Therefore, Hamilton-Santilli genoequations (3.6.66) are indeed
irreversible for all possible reversible Hamiltonians, as desired.
The origin of irreversibility rests in the contact nonpotential
forces $F^{NSA}$ according to Lagrange's and Hamilton's teaching
that is merely reformulated in an invariant way.

The above Lie-admissible mechanics requires, for completeness, {\it
three} additional formulations, the {\it backward genomechanics} for
the description of {\it matter moving backward in time,} and the
isoduals of both the forward and backward mechanics for the
description of {\it antimatter.}

The construction of these additional mechanics is lefty to the
interested reader for brevity.
\vskip0.50cm


\noindent {\bf 4. LIE-ADMISSIBLE OPERATOR MECHANICS FOR MATTER AND ITS ISODUAL FOR ANTIMATTER}

\noindent {\bf 4.1. Basic Dynamical Equations}
A simple genotopy of the naive or symplectic quantization applied to
Eqs. (3.24) yields the {\it Lie-admissible branch of hadronic
mechanics} [18] comprising four different formulations, the {\it forward and backward
genomechanics for matter and their isoduals for antimatter.} The forward
genomechanics for matter is characterized by the following main topics:

1) The nowhere singular (thus everywhere invertible) non-Hermitean {\it forward
genounit} for the representation of all effects causing
irreversibility, such as contact nonpotential interactions among extended
particles, etc. (see the subsequent chapters for various realizations)
$$
\hat I^> = 1 / \hat T^> \not = (\hat I^>)^\dag,
\eqno(4.1)
$$
with corresponding ordered product and genoreal $\hat R^>$ and genocomplex $\hat
C^>$ genofields;

2) The {\it forward genotopic Hilbert space}
$\hat {\cal H}^>$ with {\it forward genostates} $|\hat {\psi}^>>$
and {\it forward genoinner product}
$$
<^<\hat {\psi} | > |\hat {\psi}^>>\times \hat I^> =
<^<\hat {\psi} | \times \hat T^>\times |\hat {\psi}^>>\times \hat I^> \in \hat
C^>,
\eqno(4.2)
$$
and fundamental property
$$
\hat I^> > |\hat {\psi}^>> = |\hat {\psi}^>>,
\eqno (4.3)
$$
holding under the condition that $\hat I^>$ is indeed the correct unit for motion
forward in time, and {\it forward genounitary transforms}
$$
\hat U^> > (^<\hat U)^\dag{}^> = (^<\hat U)^\dag{}^> > \hat U^> = \hat I^>;
\eqno(4.4)
$$

{\sloppy 3) The fundamental Lie-admissible equations,  first proposed
in Ref.~[12] of 1974 (p. 783, Eqs. (4.18.16)) as the foundations of hadronic
mechanics, formulated on conventional spaces over conventional fields, and first
formulated in Refs. [14,18] of 1996 on genospaces and genodifferential calculus on
genofields, today's known as {\it Heisenberg-Santilli genoequations}, that  can be
written in the finite form
$$
\hat A(\hat t) = \hat U^> > \hat A(0) < ^<\hat U =  (\hat e_>^{\hat i\hat {\times}
\hat H\hat {\times}\hat t})
> \hat A(\hat 0) < (_<\hat e^{-\hat i\hat {\times}\hat t\hat {\times}
\hat H}) =
$$
$$
= (e^{i\times \hat H\times \hat T^>\times  t})\times A(0)\times
(e^{-i\times t\times ^<\hat T\times \hat H}),
\eqno(4.5)
$$
with corresponding infinitesimal version
$$
\hat i\hat {\times} {\hat d\hat A\over \hat d\hat t} = (\hat A \hat
, \hat H) = \hat A < \hat H - \hat H > \hat A =
$$
$$
= \hat A\times ^<\hat T(\hat t, \hat r, \hat p, \hat
\psi,\ldots.)\times \hat H - \hat H\times \hat T^>(\hat t, \hat r, \hat
p, \hat \psi,\ldots)\times\hat A,
\eqno(4.6)
$$
where  there is no time  arrow, since Heisenberg's
equations are computed at a fixed time.

4) The equivalent {\it Schr\"{o}dinger-Santilli genoequations}, first  suggested
in the original proposal [12] to build hadronic mechanics (see also Refs.~[17,23,24]),
formulated  via conventional mathematics and in Refs.~[14,18] via
genomathematics, that can be written
$$
\hat i^> > {\hat {\partial}^>\over \hat {\partial}^>\hat t^>}|\hat
{\psi}^>> = \hat H^> > |\hat {\psi}^>> =
$$
$$
= \hat H(\hat r, \hat v)\times \hat T^>(\hat t,\hat r, \hat p,
\hat {\psi}, \hat {\partial}\hat {\psi} \ldots)\times |\hat
{\psi}^>> = E^> > |\psi^>>,
\eqno(4.7)
$$
where the time orderings in the second term are ignored for simplicity
of notation;

}5) The {\it forward genomomentum} that escaped identification for two decades and
was finally identified thanks to the genodifferential calculus in Ref.
[14] of 1996
$$
\hat p_k^> >|\hat {\psi}^>> = - \hat i^> >\hat {\partial}_k^>|\hat
{\psi}^>> = - i\times \hat I^{>{\kern 1pt} i}_k\times
\partial_i|\hat {\psi}^>>,
\eqno(4.8)
$$

6) The {\it fundamental genocommutation rules} also first identified in Ref.~[14],
$$
(\hat r^i {\kern 1pt} \hat , {\kern 1pt} \hat
p_j) = i\times \delta^i_j\times \hat I^>,\;\; (\hat r^i {\kern 1pt} \hat
, {\kern 1pt} \hat r^j) = (\hat p_i {\kern 1pt} \hat , {\kern 1pt}
\hat p_j) = 0,
\eqno (4.9)
$$

7) The {\it genoexpectation values}  of an observable for the forward
motion $\hat A^>$ [14,19]
$$
{<^<\hat {\psi} |  > \hat A^> > |\hat {\psi}^>>\over <^<\hat {\psi}
| > |\hat {\psi}^>>}\times \hat I^> \in \hat C^>,
\eqno(4.10)
$$
under which the genoexpectation values of the genounit recovers
the conventional Planck's unit as in the isotopic case,
$$
{<\hat {\psi} |  > \hat I^> > |\hat {\psi}>\over <\hat {\psi} | >
|\hat {\psi}>} = I.
\eqno(4.11 )
$$

The following comments are now in order.
Note first in the genoaction principle  the crucial independence of isoaction
$\hat {\cal A}^>$ in
 form the linear momentum, as expressed by the Hamilton-Jacobi-Santilli
genoequations (4.3.25). Such independence assures that genoquantization
yields a genowavefunction solely dependent on time and coordinates,
$\hat \psi^> = \hat \psi^>(t, r)$.

Other geno-Hamiltonian mechanics studied previously [7] do not
verify such a condition, thus implying genowavefunctions with an
explicit dependence also on linear momenta, $\hat \psi^> = \hat
\psi^>(t, r, p)$ that violate the abstract identity of quantum and
hadronic mechanics whose treatment in any case is beyond our
operator knowledge at this writing.

Note that {\it forward geno-Hermiticity coincides with conventional
Hermiticity.} As a result, {\it all quantities that are observables for
quantum mechanics remain observables for the above genomechanics.}

However, unlike quantum mechanics, physical quantities are generally
{\it nonconserved,}  as it must be the case for the energy,
$$
\hat i^> > {\hat d^> \hat H^>\over \hat d^> \hat t^>} = \hat H\times
(^<\hat T - \hat T^>)\times \hat H \not = 0.
\eqno(4.12)
$$

Therefore, {\it the genotopic branch of hadronic mechanics is the
only known operator formulation permitting nonconserved quantities
to be Hermitean as a necessary condition to be observability.}

Other formulation attempt to represent nonconservation, e.g., by
add\-ing an ``imaginary potential'' to the Hamiltonian, as it is often done in
nuclear physics [25]. In this case the Hamiltonian is non-Hermitean and,
consequently, the nonconservation of the energy cannot be an observable.

Besides, said ``nonconservative models'' with non-Hermitean
Hamiltonians are nonunitary and are formulated on conventional
spaces over conventional fields, thus suffering all the catastrophic
inconsistencies of Theorem 1.3.

We should stress the representation of irreversibility and nonconservation
beginning with the most primitive quantity, the unit and related product.
{\it Closed irreversible systems} are characterized by the Lie-isotopic
subcase in which
$$
\hat i\hat {\times} {\hat d\hat A\over \hat d\hat t} = [\hat A \hat
, \hat H] = \hat A\times \hat T(t, \ldots)\times \hat H - \hat H\times
\hat T(t, \ldots)\times\hat A,
\eqno4.13a)
$$
$$
^<{\hat T}(t, \ldots) = \hat T^>(t, \ldots) = \hat T(t,
\ldots) = \hat T^\dag(t, \ldots) \not = \hat T(-t, \ldots),
\eqno(4.13b)
$$
for which the Hamiltonian is manifestly conserved. Nevertheless the
system is manifestly irreversible. Note also the first and
only known observability of the Hamiltonian (due
to its iso-Hermiticity) under irreversibility.

As one can see, brackets $(A, B)$ of Eqs. (4.6) are jointly
Lie- and Jordan-admissible.

Note also that finite genotransforms (4.4.5) verify the condition of
genohermiticity, Eq. (4.4).

We should finally mention that, as it was the case for isotheories, {\it
genotheories are also admitted by the abstract axioms of quantum mechanics, thus
providing a broader realization.} This can be seen, e.g., from the invariance
under a complex number $C$
$$
<\psi| x |\psi> \times I = <\psi| x C^{-1} \times  |\psi> \times (C\times I) =
<\psi| > |\psi> \times I^>.
\eqno(4.14)
$$

Consequently, {\it genomechanics provide another explicit and concrete
realization of ``hidden variables'' [26], thus constituting another ``completion''
of quantum mechanics in the E-P-R sense [27].} For the studies of these aspects we
refer the interested reader to Ref. [28].

 The above formulation must be completed with three additional
Lie-admis\-sible formulations, the backward formulation for matter
under time reversal and the two additional isodual formulations for
antimatter. Their study is left to the interested reader for
brevity.
\vskip0.30cm


\noindent {\bf 4.2. Simple Construction of Lie-Admissible Theories}
As it was the case for the isotopies, a simple method has been
identified in Ref. [44] for the construction of Lie-admissible
(geno-) theories from any given conventional, classical or quantum
formulation. It consists in {\it identifying the genounits as the
product of  two different nonunitary transforms,}
$$
\hat I^> = (^<\hat I)^\dag = U\times W^{\dagger}, \; \; \; ^<{\hat I}
= W\times U^\dag,
\eqno(4.15a)
$$
$$
U\times U^{\dagger} \not = 1,\;\;\; W\times W^{\dagger} \not = 1,\;\;\; U\times
W^{\dagger} = \hat I^>,
\eqno(4.15b)
$$
and subjecting the totality of quantities and their operations of
conventional models to said dual transforms,
$$
I\rightarrow \hat I^> = U\times I\times W^{\dagger},\;\;\; I\rightarrow
^<\hat I = W\times I\times U^{\dagger},
\eqno(4.16a)
$$
$$
a\rightarrow \hat a^> = U\times a\times W^{\dagger} = a\times \hat
I^>,
\eqno(4.16b)
$$
$$
 a\rightarrow
^<\hat a = W\times a\times U^{\dagger} = ^<\hat I\times a,
\eqno(4.16c)
$$
$$
a\times b\rightarrow \hat a^> > \hat b^> = U\times (a\times b)\times
W^> =
$$
$$
= (U\times a\times W^{\dagger})\times (U\times
W^{\dagger})^{-1}\times (U\times b\times W^{\dagger}),
\eqno(4.16d)
$$
$$
\partial/\partial x\rightarrow \hat {\partial}^>/\hat {\partial}^>\hat
x^> = U\times (\partial/\partial x)\times W^\dagger = \hat
I^>\times(\partial/\partial x),
\eqno(4.16e)
$$
$$<\psi|\times |\psi>\rightarrow <^<\psi| >
|\psi^>> = U\times (<\psi|\times |\psi >)\times W^{\dagger},
\eqno(4.16f)
$$
$$
H\times |\psi >\rightarrow \hat H^> > |\psi^>> =
$$
$$
= (U\times H\times W^{\dagger})\times (U\times
W^{\dagger})^{-1}\times (U\times \psi>W^\dag), \ {\rm etc.} \eqno(4.4.16g)
$$

As a result, any given conventional, classical or quantum model can
be easily lifted into the  genotopic form.

Note that the above construction implies that {\it all conventional
physical quantities acquire a well defined direction of time.} For
instance, the correct genotopic formulation of  energy, linear
momentum, etc., is given by
$$
\hat H^> = U\times H\times W^\dag,\;\; \hat p^> = U\times p\times W^>, \
{\rm etc.}
\eqno(4.17)
$$
In fact, under irreversibility, the value of a nonconserved energy
at a given time $t$ for motion forward in time is generally
different than the corresponding value of the energy for $-t$ for
motion backward in past times.

This explains the reason for having represented in this section
energy, momentum and other quantities with their arrow of time $>$.
Such an arrow can indeed be omitted for notational simplicity, but
only after the understanding of its existence.

Note finally that  a conventional, one dimensional, unitary Lie transformation
group with Hermitean generator $X$ and parameter $w$ can be transformed into a
covering Lie-admissible group via the following nonunitary transform
$$
Q(w)\times Q^\dag(w) = Q^\dag(w)\times Q(w) = I,\; w \in R,
\eqno(4.18a)
$$
$$
U\times U^\dag \not = I, \; \; \; W\times W^\dag \not = 1,
\eqno(4.18b)
$$
$$
A(w) = Q(w)\times A(0) \times Q^\dag(w) =  e^{X\times w\times i} \times A(0) \times
e^{- i\times w\times X} \rightarrow
$$
$$
\rightarrow U\times (e^{X\times w\times i} \times A(0) \times
e^{- i\times w\times X})\times U^\dag =
$$
$$
\equiv [U\times (e^{X\times w\times i})\times W^\dag \times (U\times
W^\dag)^{-1}\times A\times A(0)\times
$$
$$
\times  U^\dag \times (W\times U^\dag)^{-1}\times
[W\times  (e^{- i\times w\times X})\times U^\dag] =
$$
$$
= (e^{i\times X\times X})^> > A(0) < ^<(e^{-1\times w\times X}) = \hat U^> > A(0) <
^<\hat U,
\eqno(4.18c)
$$
which confirm the property of Section 4.2, namely, that under the necessary
mathematics {\it the Lie-admissible theory is indeed admitted by the abstract Lie
axioms, and it is a realization of the latter broader than the isotopic form.}

\vskip0.30cm


\noindent {4.3. Invariance of Lie-Admissible Theories}
Recall that a fundamental axiomatic feature of quantum mechanics is
the invariance under time evolution of all numerical predictions and
physical laws, which invariance is due to the {\it unitary
structure} of the theory.

However, quantum mechanics is reversible and can only represent in
a scientific way beyond academic beliefs reversible systems verifying
total conservation laws due to the antisymmetric character of the
brackets of the time evolution.

As indicated earlier, the representation of irreversibility and
nonconservation requires theories with a {\it nonunitary structure}.
However, the latter are afflicted by the catastrophic
inconsistencies of Theorem 1.3.

The only resolution of such a basic impasse known to the author has
been the achievement of invariance under nonunitarity and
irreversibility via the use of genomathematics, provided that such
genomathematics is applied to the {\it totality} of the formalism to
avoid evident inconsistencies caused by mixing different mathematics
for the selected physical problem.\

Let us nmote that, due to decades of
protracted use it is easy to predict that physicists and
mathematicians may be tempted to treat the Lie-admissible branch of
hadronic mechanics with conventional mathematics, whether in part or
in full. Such a posture would be equivalent, for instance, to
the elaboration of the spectral emission of the hydrogen atom with
the genodifferential calculus, resulting in an evident nonscientific
setting.

Such an invariance was first achieved by Santilli in Ref. [15] of
1997 and can be illustrated by reformulating any given nonunitary
transform in the {\it genounitary form}
$$
U = \hat U\times \hat T^{>1/2}, W = \hat W\times \hat T^{>1/2},
\eqno(4.19a)
$$
$$
U\times W^{\dagger} = \hat U > \hat W^{\dagger} = \hat W^{\dagger} >
\hat U = \hat I^> = 1/\hat T^>,
\eqno(4.19b)
$$
and then showing that genounits, genoproducts, genoexponentiation,
etc., are indeed invariant under the above genounitary transform in
exactly the same way as conventional units, products,
exponentiations, etc. are invariant under unitary transforms,
$$
\hat I^>\rightarrow \hat I^{>'} = \hat U > \hat I^> > \hat
W^{\dagger} = \hat I^>,
\eqno(4.20a)
$$
$$
\hat A > \hat B\rightarrow \hat U > (A > B) > \hat W^{\dagger} =
$$
$$
= (\hat U\times \hat T^>\times A\times T^>\times \hat
W^{\dagger})\times (\hat T^>\times W^{\dagger})^{-1}\times \hat
T^>\times
$$
$$
\times (\hat U\times \hat T^>)^{-1}\times (\hat U\times T^>\times
\hat A\times T^>\times \hat W^>) =
$$
$$
= \hat A'\times (\hat U\times \hat
W^{\dagger})^{-1}\times \hat B = \hat A'\times \hat T^>\times B' =
\hat A' > \hat B', \ {\rm etc.}
\eqno(4.20b)
$$
from which all remaining invariances follow, thus resolving the
catastrophic inconsistencies of Theorem 1.3.

Note the {\it numerical invariances of the genounit} $\hat
I^>\rightarrow \hat I^{>'} \equiv \hat I^>$, {\it of the genotopic
element}  $\hat T^>\rightarrow \hat T^{>'} \equiv \hat T^>$, {\it
and of the genoproduct} $>\rightarrow >' \equiv >$ that are
necessary to have invariant numerical predictions.

\vskip0.50cm

\noindent {\bf 5. SOCIETAL AND SCIENTIFIC IMPLICATIONS}

\noindent The societal, let alone scientific implications of the proper treatment of irreversibility are
rather serious. Our planet is afflicted by increasingly catastrophic climactic events mandating the
search for basically new, environmentally acceptable energies and fuels.

All known energy sources, from the combustion of carbon dating to prehistoric times to the contemporary
nuclear energy, are based on structurally irreversible processes. By comparison, all established
doctrines  of the 20-th century, such as quantum mechanics and special relativity, are structurally
reversible, that is, reversible in their basic axioms, let alone their physical laws.

It is then easy to see that {\it the serious search for basically new energies and fuels will require
basically new theories that are as structurally irreversible as the process they are expected to describe.}
At any rate, all possible energies and fuels that could be predicted by quantum mechanics and special
relativity were discovered by the middle of the 20-th century. Hence, the insistence in
continuing to restrict irreversible processes to comply with preferred reversible
doctrines may perhaps yield myopic short terms benefits, but also cause a potentially
historical condemnation by posterity.

An effective way to illustrate the need for new irreversible theories is given by nuclear fusions. All
efforts to date in the field, whether for the ''cold fusion'' or the ''hot fusion,'' have been studiously
restricted to verify special relativity and relativistic quantum mechanics. However, {\it whether ''hot'' or
''cold,'' all fusion processes are strictly irreversible, while special relativity and relativistic quantum
mechanics are strictly reversible.}

It has been shown in Ref. [73] that some of the reasons for the
failure to date by both the ''cold'' and the ''hot'' fusions to achieve industrial value is due precisely
to the treatment of structurally irreversible nuclear fusions with structurally reversible mathematical and
physical methods.

In the event of residual doubt due to protracted use of preferred theories, it is
sufficient to compute  the quantum mechanical probability for two nuclei to ''fuse'' into a third
one, and then compute its time reversal image. In this way the serious scholar will see that special
relativity and relativistic quantum mechanics predict a fully causal {\it spontaneous disintegration of
nuclei following their fusion,} namely, a prediction outside the boundary of serious science.

The inclusion of irreversibility in quantitative studies of new energies then recommend the
development, already partially achieved at the industrial level (see Chapter 8 of Ref. [19]), of a new,
controlled ''intermediate fusion'' of light nuclei [73], that is, a fusion occurring at minimal
energies firstly needed to expose nuclei as a pre-requisite for their fusion (a feature absent in the
''cold fusion'' due to insufficient energies), and  secondly necessary to prevent instabilities at the
very high energies of the ''hot fusion'' that have been uncontrollable precisely because due to
irreversible processes described by reversible doctrines.

In view of these pressing societal needs, the construction of Lie-admissible and/or Lie-isotopic coverings of
special relativity and relativistic quantum mechanics is advocated, followed by experimental
verifications and specific applications to new clean energies so much needed by mankind.

Recall that Lorentz,
Poincar\'e, Einstein, Minkowski and others insisted in the reversible character of special relativity
because necessary for the description of the physical events for which the relativity was built for.

It is easy to see that, following the above historical teaching, a basically new relativity must be
developed for irreversible processes. In fact, the original proposal of a Lie-admissible generalization of
Lie's theory of 1978 [11] was intended, specifically, for the construction of a Lie-admissible covering of
Galileo relativity indicated beginning with the title.  The proposed relativity is expected to be a
genotopic covering of the {\it isorelativity} already  constructed and verified
[18b,18c].

Another illustration of the implications of the studies herein considered is given by the fact that   {\it
all inelastic scatterings  in particle physics are irreversible, yet they have been elaborated in the 20-th
century via the conventional  (potential) scattering theory which is structurally reversible.} Once
possible contributions from irreversibility are duly taken into account, some of the ''experimental
results'' in inelastic scatterings of the 20-th century may well turn out to be ''experimental beliefs.''

In order to implement a serious scientific process, rather than follow a scientific religion, it is
necessary to construct {\it Lie-admissible (Lie-isotopic) scattering theory for open (closed) inelastic
reactions}  via the procedure given in  Section 4.2, re-examine existing results on inelastic scatterings
and establish whether or not irreversibility requires a revision of existing data. In particular, the
proposed irreversible scattering  theory is expected to be an extension of the isotopic scattering theory
already embrionically constructed (see Ref. [18b] and contributions quoted therein).

Far from being inessential, the re-examination of
irreversible particle processes via an axiomatically consistent and invariant irreversible mechanics
appears to require a revision of hadron physics for which hadronic mechanics was built for
[12,18]. In fact, various studies (see, e.g., Refs. [74,75] and literature quoted therein) have shown the
need for a serious re-examination of  quarks, neutrinos and other conjectures, again, to prevent that
possible myopic short term gains in reality set the foundations for a potentially historical condemnation by
posterity.

After all, quarks and neutrino cannot be detected directly; their existence is claimed only on grounds of
conventional particles predicted from inelastic scattering elaborated via a reversible scattering theory;
and the resulting events admits numerous alternative interpretations, including those via photons emitted
by antiparticles. In any case, quarks are known not to admit gravity (because they can only be defined in a
mathematics internal space while gravity can only be defined in our spacetime), and admit a rather long list
of additional basically unsolved insufficiencies [18c,73-75].

In closing, the main objectives of this paper are to establish beyond credible doubt that the problem of
the origin of irreversibility is more open then ever; to illustrate the societal, let alone scientific need
for its systematic study; to propose a quantitative, axiomatically consistent representation of
irreversibility at the classical and operator levels as well as for matter and antimatter; and to solicit
alternative formulations by interested colleagues, under the condition that they are as directly universal
and invariant as the proposed Lie-admissible and Lie-isotopic formulations.

Finally, a few words on the {\it limitations} of our Lie-admissible and Lie-isotopic treatments of
irreversibility are in order. We have stressed in Section 1 that physics will never admit final theories.
That is the fate also for our  formulations. In fact, their most visible limitations are due
to the fact that {\it the basic genounits, their genoproducts and other operations are ''single-valued,''}
namely, the result of genooperations are given by one single quantity.

Such a feature is today known to be effective for irreversible classical and operator {\it physical
processes,} but said feature is also known to be insufficient for the representation of {\it
biological structure,} since the latter require not only clearly irreversible methods but also a {\it
multi-valued generalization of the (single-valued) Lie-admissible and Lie-isotopic formulations,} whose
construction has been already embrionically initiated via the {\it hyperstructural branch of hadronic
mechanics} [18-22].

It is hoped that serious scholars will participate with independent studies on the
above, as well as numerous other, basically open problems because, in the final analysis,  lack of
participation in basic advances is a gift of scientific priorities to others.

\vskip0.30cm

\noindent {\large \bf Acknowledgments}

\noindent The author would like to thank for invaluable critical comments  all
participants of the {\it XVIII Workshop on Hadronic Mechanics} held at the
University of Karlstad, Sweden, on  June 20-22, 2005.

\end{document}